\documentclass[bibtex]{pasj02} 
\usepackage[switch,mathlines]{lineno} % add line number to manuscript
\usepackage{natbib} 
\usepackage{hyperref}

\jyear{2025}
\Received{}
\Accepted{}

\newcommand{\kms}{${\rm km\,s^{-1}}$}
\newcommand{\netunit}{cm$^{-3}$s}
\newcommand{\msun}{$M_{\rm \odot}$}
\newcommand{\chandra}{{Chandra}}

\newcommand{\xrism}{{XRISM}}
\newcommand{\resolve}{{Resolve}}
\newcommand{\xspec}{{xspec}}
\newcommand{\sect}{{section}}

\begin{document} 

\title{Mapping Cassiopeia A's silicon/sulfur Doppler velocities with XRISM-Resolve}

\author{
Jacco \textsc{Vink},\altaffilmark{1,2}\orcid{0000-0002-4708-4219}
\email{j.vink@uva.nl}
Manan \textsc{Agarwal},\altaffilmark{1}\orcid{0000-0001-6965-8642}
\email{m.agarwal@uva.nl}
Aya \textsc{Bamba},\altaffilmark{3,4,5}\orcid{0000-0003-0890-4920}
Liyi \textsc{Gu},\altaffilmark{2}\orcid{0000-0001-9911-7038}
Paul \textsc{Plucinsky},\altaffilmark{6}\orcid{0000-0003-1415-5823}
Ehud \textsc{Behar},\altaffilmark{7}\orcid{0000-0001-9735-4873}
Lia \textsc{Corrales},\altaffilmark{8}\orcid{0000-0002-5466-3817} 
Adam \textsc{Foster},\altaffilmark{6}\orcid{0000-0003-3462-8886}
Shin-ichiro \textsc{Fujimoto},\altaffilmark{9}\orcid{0000-0002-7273-2740}
Masahiro \textsc{Ichihashi},\altaffilmark{3}\orcid{0000-0001-7713-5016}
Kazuhiro \textsc{Ichikawa},\altaffilmark{10}\orcid{0009-0009-6901-7955}
Satoru \textsc{Katsuda},\altaffilmark{11}\orcid{0000-0002-1104-7205} 
Hironori \textsc{Matsumoto},\altaffilmark{12} 
Kai \textsc{Matsunaga},\altaffilmark{13}\orcid{0009-0003-0653-2913} 
Tsunefumi \textsc{Mizuno},\altaffilmark{14}\orcid{0000-0001-7263-0296}
Koji \textsc{Mori},\altaffilmark{10}\orcid{0000-0002-0018-0369}
Hiroshi \textsc{Murakami},\altaffilmark{15}\orcid{0000-0002-3844-5326}
Hiroshi \textsc{Nakajima},\altaffilmark{17}\orcid{0000-0001-6988-3938}
Toshiki \textsc{Sato},\altaffilmark{18}\orcid{0000-0001-9267-1693}
Makoto \textsc{Sawada},\altaffilmark{19}\orcid{0000-0003-2008-6887}
Haruto \textsc{Sonoda},\altaffilmark{20} 
Shunsuke \textsc{Suzuki},\altaffilmark{20} 
Dai \textsc{Tateishi},\altaffilmark{3,21}\orcid{0000-0003-0248-4064}
Yukikatsu \textsc{Terada},\altaffilmark{11,20}\orcid{0000-0002-2359-1857}
Hiroyuki \textsc{Uchida}\altaffilmark{13}\orcid{0000-0003-1518-2188}
}

\altaffiltext{1}{Anton Pannekoek Institute/GRAPPA, University of Amsterdam, Science Park 904, 1098 XH Amsterdam, The Netherlands}
\altaffiltext{2}{SRON Netherlands Institute for Space Research, Niels Bohrweg 4, 2333 CA Leiden, The Netherlands}
\altaffiltext{3}{Department of Physics, Graduate School of Science,
The University of Tokyo, 7-3-1 Hongo, Bunkyo-ku, Tokyo 113-0033, Japan}
\altaffiltext{4}{Research Center for the Early Universe, School of Science, The University of Tokyo, 7-3-1
Hongo, Bunkyo-ku, Tokyo 113-0033, Japan}
\altaffiltext{5}{Trans-Scale Quantum Science Institute, The University of Tokyo, Tokyo  113-0033, Japan}
\altaffiltext{6}{Harvard-Smithsonian Center for Astrophysics, MS-3, 60 Garden Street, Cambridge, MA, 02138, USA}
\altaffiltext{7}{Department of Physics, Technion, Technion City, Haifa 3200003, Israel}
\altaffiltext{8}{Department of Astronomy, University of Michigan, MI 48109, USA}
\altaffiltext{9}{National Institute of Technology Kumamoto College, 2659-2 Suya, Koshi, Kumamoto 861-1102, Japan}
\altaffiltext{10}{Faculty of Engineering, University of Miyazaki, Miyazaki 889-2192, Japan}
\altaffiltext{11}{Graduate School of Science and Engineering, Saitama University, 255 Shimo-Ohkubo, Sakura, Saitama 338-8570, Japan}
\altaffiltext{12}{Department of Earth and Space Science, Osaka University, Osaka 560-0043, Japan}
\altaffiltext{13}{Department of Physics, Kyoto University, Kyoto 606-8502, Japan}
\altaffiltext{14}{Department of Physics, Hiroshima University, Hiroshima 739-8526, Japan}
\altaffiltext{15}{Department of Data Science, Tohoku Gakuin University, Miyagi 984-8588, Japan}
\altaffiltext{17}{College of Science and Engineering, Kanto Gakuin University, Kanagawa 236-8501, Japan}
\altaffiltext{18}{School of Science and Technology, Meiji University, Kanagawa, 214-8571, Japan}
\altaffiltext{19}{Department of Physics, Rikkyo University, Tokyo 171-8501, Japan}
\altaffiltext{20}{ISAS/JAXA, 3-1-1 Yoshinodai, Chuo-ku, Sagamihara, Kanagawa 252-5210, Japan}
\altaffiltext{21}{
Present address: Collaborative Laboratories for Advanced Decommissioning Science, Japan Atomic Energy Agency, Fukushima, Japan
}


\maketitle

\begin{abstract}
Young supernova remnants (SNRs) provide crucial insights into explosive nucleosynthesis products and their velocity distribution soon after the explosion. However, these velocities are influenced by the dynamics of the circumstellar medium (CSM), which in young core-collapse SNRs originates from the progenitor's late-phase mass loss. 
Cassiopeia A (Cas A), the youngest known Galactic core-collapse SNR, was studied to analyze the spatial distribution of silicon and sulfur radial velocities using two high-spectral resolution observations from the \xrism\ \resolve\ imaging spectrometer. \resolve's capabilities enabled the detailed characterization of Si XIII, Si XIV, S XV, and S XVI lines, whose line shapes can be resolved and  modeled using Gaussian radial-velocity components.
The radial velocities measured generally align with previous CCD-based results, confirming that they were not artifacts caused by blended lines or ionization variations. Modeling line profiles with two-component Gaussians improved fits in some regions, revealing distinct redshifted (backside) and blueshifted (frontside) components only in a few specific areas. In most regions, however, both components were either both redshifted (northwest) or both blueshifted (southeast), consistent with the patchy ejecta shell morphology seen in optically emitting fast-moving knots. 
The individual line components revealed a
line broadening ranging from $\sigma_v \approx 200$ to $\sigma_v \approx 2000$~\kms. Components with $1000 \lesssim \sigma_v \lesssim 2000$~\kms\ are consistent with previously determined reverse shock velocities, suggesting non-equilibrated or partially equilibrated ion temperatures.
Narrow components with small radial velocities found
near Cas A's projected center likely originate from shocked CSM plasma. 
But the low radial velocity and small $\sigma_v$ defies identifying
these components with either the frontside or backside of the SNR, or both.

\end{abstract}

%\pagewiselinenumbers 
 \nolinenumbers

\section{Introduction}

% flatex input: [casa_velocities_fig_pixels.tex]
\begin{figure}
\includegraphics[width=\columnwidth]{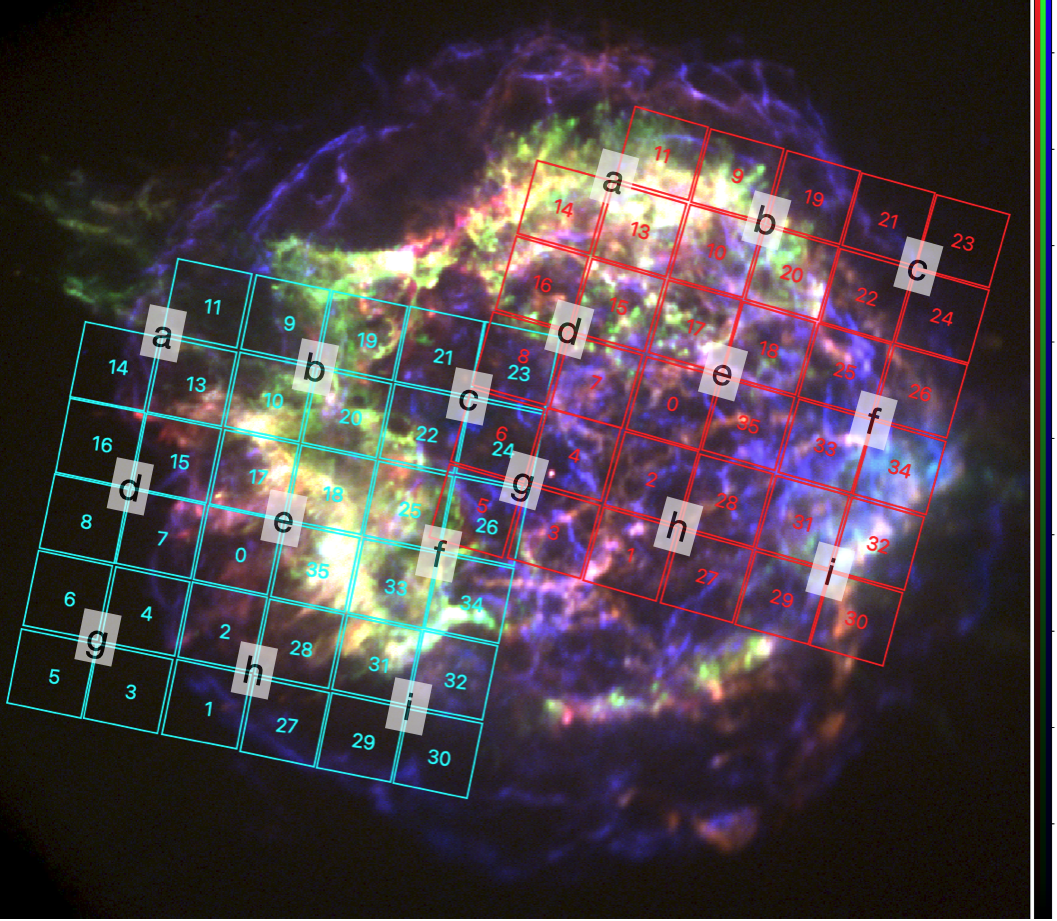}
 \caption{
 XRISM Resolve pixel numbering projected onto a three-color Chandra image of Cas A, using the
 2004 observations \citep{hwang04}. Red: 1.28--1.4 keV; green: 1.75--1.95 keV; blue: 4-6 keV.
 Apart from the pixel numbers we also show 2$\times$2 pixel binnings, labeled
 per observation from {\it a} to {\it i}.
{Alt text: A three-color image of the supernova remnant Cas A in X-rays, as
observed by Chandra. The two Resolve field-of-views are overlaid, with pixel numbers.}
 \label{fig:pixels}
 }
 \end{figure}

\begin{figure*}
  \centerline{
    \includegraphics[trim=0 0 0 0,clip=true,width=0.5\textwidth]{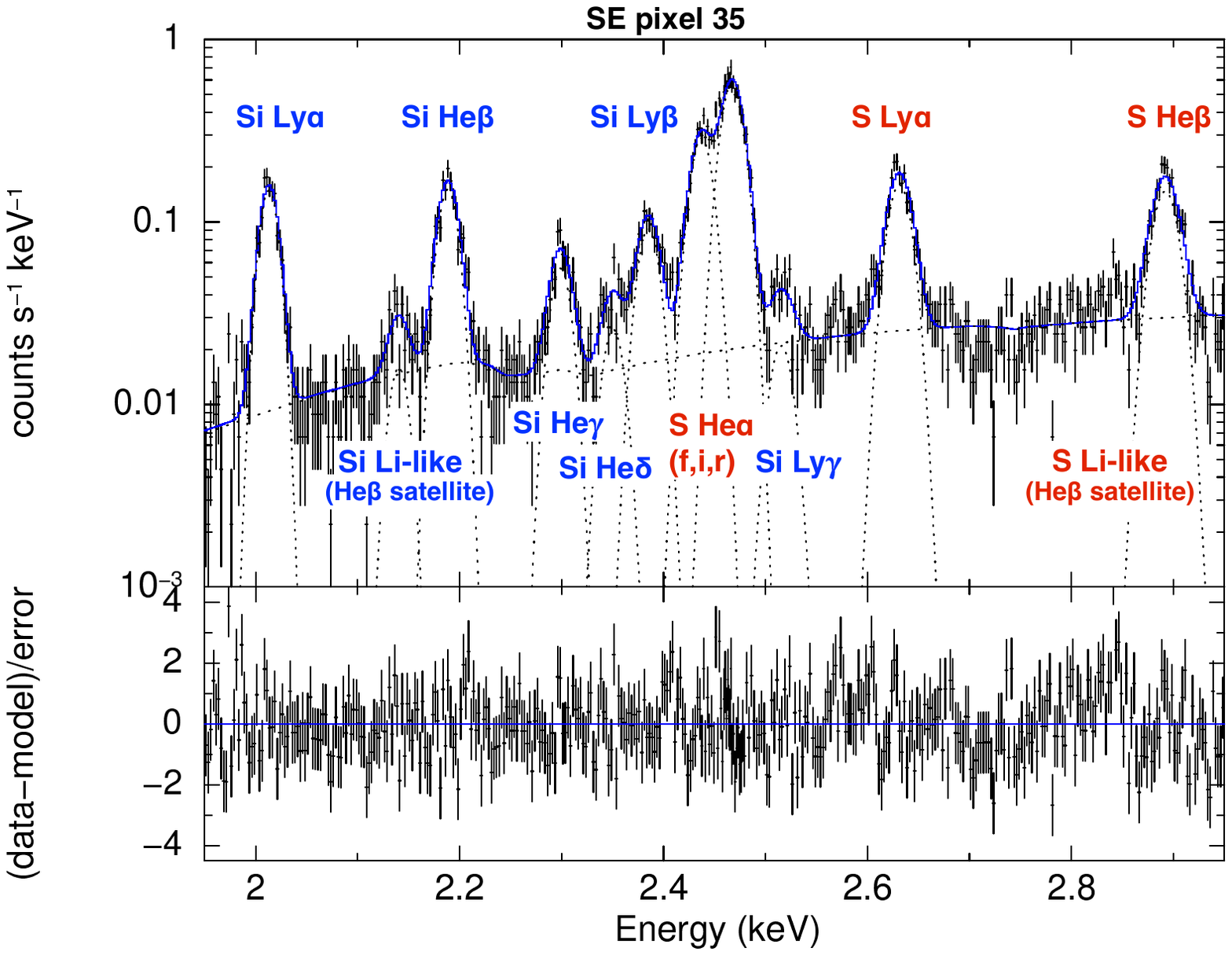}    
    \includegraphics[trim=0 0 0 0,clip=true,width=0.5\textwidth]{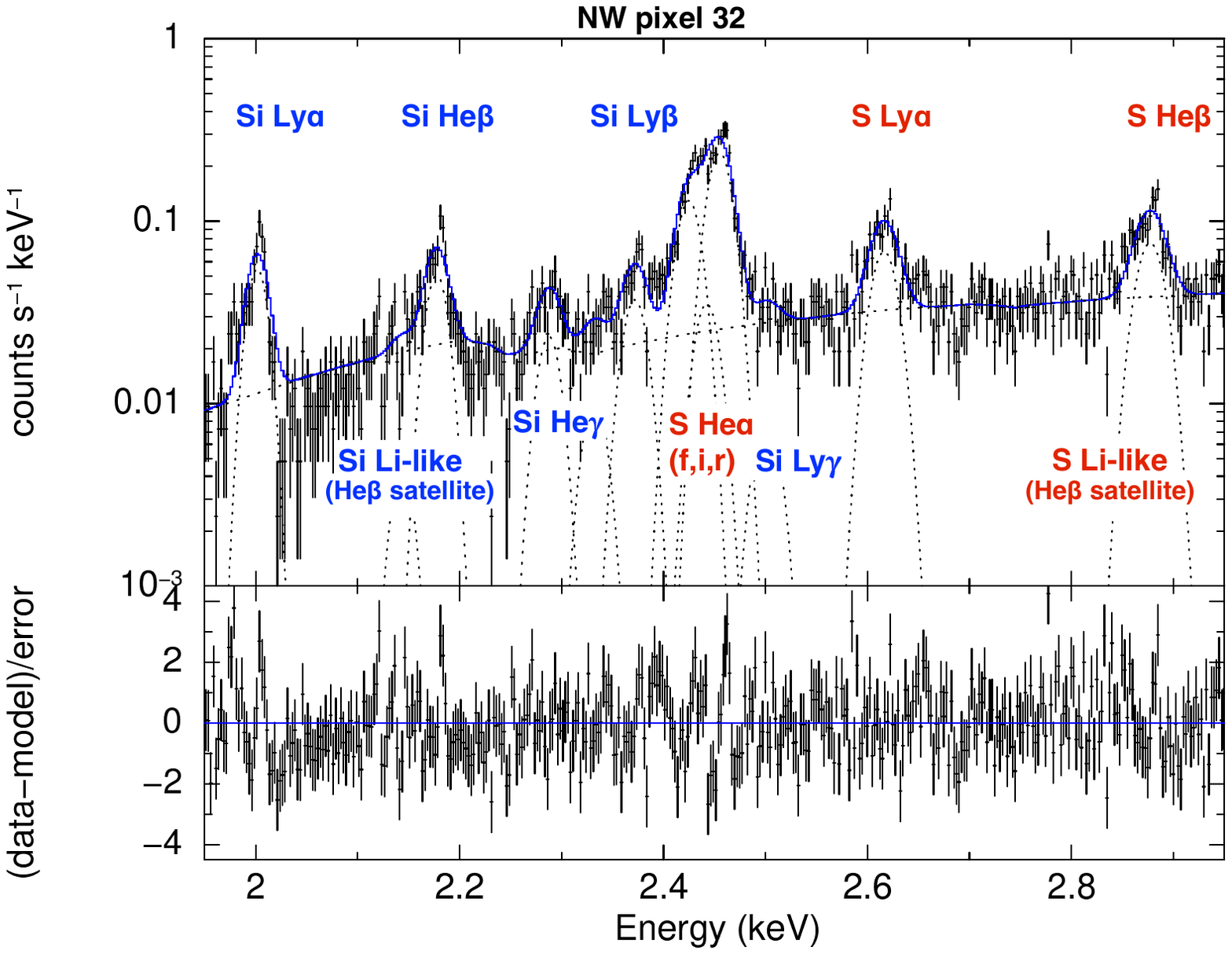}
    }
  \caption{
    Two examples of the spectral fits obtained with the automated spectral
    fitting method. The individual spectral components are indicated with
    dotted lines. Left: SE region pixel 35. Right: NW region pixel 32.
     Line identifications are given.
    {Alt text: Two panels with in each a Resolve X-ray spectrum from 1.9-2.9 keV, and the best-fit model using a continuous line. The lower half shows the fit residuals. Left a SE spectrum (pixel 35), right a NW spectrum (pixel 32).
    } 
    \label{fig:spectra}
  }
\end{figure*}

The dynamics of  supernova remnants (SNRs), both those of the blast wave and reverse shock,
 are determined  by the explosion properties and by the density distribution of the circumstellar medium (CSM) encountered  by the blast wave.
The spatial and velocity distributions of fresh nucleosynthesis products  are indicative of the velocities imparted on them by the supernova explosion itself, but is also shaped by
 the reverse shock, which heats the ejecta, revealing their presence through X-ray radiation.  
For core-collapse SNRs the interest in the dynamics concerns both the role of the last phases of stellar mass loss, which shape the CSM, as well as the asymmetries in the ejecta velocities imparted by the
explosion. 
Asymmetries in ejecta velocities can be caused by the interplay between matter infalling onto the proto-neutron stars, and the energetic neutrinos emitted by the proto-neutron star
in the first few seconds following core collapse. The explosion may be aided, or even dominated, by magnetorotational instabilities generated around the rapidly rotating magnetized proto-neutron stars, resulting in
jet formation \citep[e.g.][]{burrows07,moesta15}. 

One of the best objects to study both the explosion dynamics and the structure of the CSM is the youngest-known Galactic core-collapse SNR Cassiopeia A (Cas A). Cas A is $\sim 353$~yr old \citep{thorstensen01}
and located at a distance of 3.4~kpc \citep{reed95}. Its  outer blast wave, marked by X-ray synchrotron radiation \citep[e.g.][]{vink04a}, is surprisingly circular. In contrast, Cas A's shocked ejecta 
form a highly irregular shell, both in the infrared/optical---in the form of numerous dense ejecta knots \citep[e.g.][]{alarie14,milisavljevic24}---and in X-rays \citep{hwang12}. 
The X-ray spectra are characterized by bright line emission from intermediate mass elements (IMEs), in particular line emission from helium-like Si, S, and Ar, and in addition Fe-L and Fe-K line emission.
The Fe emission can be found throughout the SNR shell, but in the SE there are protrusions of iron-rich plasma, indicating that some Fe-rich ejecta
was ejected with  at least $\sim 7000$~\kms\ \citep{hughes01,hwang12}.
There is one striking feature seen in both the optical \citep[e.g.][]{fesen06}  and X-rays 
\citep{vink04a,hwang04} suggesting some axis of  symmetry in the explosion: a bipolar structure rich in IMEs, the so-called Cas A ``jets".

Cas A 
likely exploded after having shed most of its outer hydrogen envelope, leaving only 2--4~\msun\ of material to eject \citep{vink96,willingale03,hwang12}. The progenitor, therefore, consisted of
not much more than the helium-core of what was once a 16---20\msun\ star \citep{chevalier03}.
Indeed, an optical light echo spectrum identifies the supernova as being of Type IIb---i.e. a stripped envelope supernova \citep{krause08,rest11}. 

The three-dimensional structure  of the optical knots has been well studied using optical spectroscopy \citep[e.g.][]{reed95,lawrence95,alarie14}. For a long time X-ray astronomy lacked
the means of high-spectral resolution imaging spectroscopy.  Early X-ray evidence for  clear asymmetry in Doppler shifts from the northwest to southeast 
was obtained using the Focal Plane Crystal Spectrometer (FPCS) on board the Einstein telescope \citep{markert83}. This Doppler velocity asymmetry was later confirmed with better spatial resolution using CCD spectroscopy \citep{holt94,willingale02,hwang01b}. X-ray CCD detectors have only a modest spectral resolution of $E/\Delta E \sim 20$--$50$, corresponding to $\sigma_v\sim  15,000$--$6000~{\rm km\ s^{-1}}$.
The measured Doppler velocity asymmetries---from $\sim -2000~{\rm km\,s^{-1}}$ to  $\sim 2000~{\rm km\,s^{-1}}$---were smaller than the energy resolution as measurements relied on  
centroiding, specifically
the bright Si XIII He$\alpha$,  S XV He$\alpha$,  and Fe K line complexes  \citep[e.g][]{willingale02}. 
These Doppler velocity measurements are, therefore, sensitive to
changes in the line centroids caused by other effects than
Doppler shifts. In particular, the He$\alpha$ emission consists of three lines, whereas for  the Fe K line complex not only He-like iron, but also lower charge states contribute.

Both the Chandra X-ray Observatory and XMM-Newton also carry grating spectrometers, but as these are slitless the analysis of extended sources is complicated. The Chandra High-Resolution Transmitting Gratings (HETGS)
were used to measure Doppler shifts of 17 bright silicon-rich knots in Cas A, which showed that Doppler velocities ranged from $\sim -2500~{\rm km\,s^{-1}}$ to  $\sim 4000~{\rm km\,s^{-1}}$ \citep{lazendic06}.
Most of the blueshifted knots appear to be located in the SE, whereas most of the redshifted knots appear in the N/NW part of Cas A, but with some exceptions, in particular in the N/NW.
The analysis of Cas A with the HETGS is complex, and in particular taking into account the detailed spatial shape of the knots gives rise to systematic uncertainties, as exemplified in different line shapes measured using the positive and negative dispersion directions \citep{lazendic06}.
 
 With the launch of  the X-Ray Imaging and Spectroscopy Mission \citep[XRISM,][]{tashiro24} on September 6, 2023 a new high-spectral resolution spectrometer, \resolve\  \citep{ishisaki22} 
 has become available for high-spectral resolution imaging spectrometry. \resolve\ has 
 an energy resolution of $\Delta E\approx 5$~eV and resolving power of $E/\Delta E\approx 500$ at 2.5 keV and a field of view of 3.1\arcmin$\times$ 3.1\arcmin,
 divided in 35 pixels (one corner pixel is not filled). The angular resolution  of \resolve\ is modest:  1.27\arcmin\ half-power diameter (HPD). \resolve\ finally allows for measuring not only Doppler shifts of the line
 emission in Cas A, 
 but also the shapes of the lines.

\section{Data and reduction}\label{sec:data}

Cas A was observed by \xrism\ on from December 11 to 17, 2023, with one pointing targeting
the southeast region (``SE", ObsID 000129000, $\sim 181$~ks) and one targeting the northwest region (``NW", ObsID 000130000, $\sim 166$~ks).
The field of view (FoV) of \resolve\ for both observations is shown in Fig.\ref{fig:pixels} overlaid on a Chandra X-ray image. 
It shows the layout of the two 35 pixel arrays covering part of Cas A.
The data were reduced and calibrated  with the HEASoft 6.34 software package and XRISM CalDB 9 (Version 20240815).
For a proper analysis it is important to extract spectra per individual pixel, which was done using only the
High-resolution Primary (HP)
event grades, ensuring the highest spectral resolution.
For these observations the energy gain was tracked in high-fiducial calibration mode, allowing also for accurate gain calibration of the erratically behaving pixel 27. The overall gain accuracy is of the order 0.01~eV.

Each pixel has an angular extent of 30\arcsec $\times$ 30\arcsec, which oversamples the X-ray telescope's point spread function (PSF).
So it should be kept in mind that there is considerable angular correlations in  a study based on individual pixels. See \citet{plucinsky25} for a quantitative evaluation of angular correlations for these Cas A observations, and other systematic effects for these observations. 
For that reason, and for computational reasons as discussed below, we
also introduced ``super pixels" of 60\arcsec $\times$ 60\arcsec, which we labeled {\it a} to {\it i}, also shown in Fig.~\ref{fig:pixels}.

Note that the PSF effects are position dependent,
because for intrinsically bright regions next to a fainter
region a larger fraction
of the photons detected in a pixel will come from the corresponding sky region itself. The reverse is true for faint sky regions bordering brighter sky regions.
For the SE pointing, the fraction of photons coming intrinsically from that region of the sky corresponding to a superpixel ranges from $\sim 9\%$ (pixel {\it g}) to $\sim 48\%$ (pixel {\it e}).
For the NW pointing, this range is 16\% (pixel {\it c}) to $\sim 53\%$ (pixel {\it i}). 
As will be appreciated, the PSF effects are especially severe near the edges of Cas A, i.e. the SE corner and NW corner of Cas A.
More details are provided in \citet{plucinsky25}.

\begin{figure*}
  \centerline{
    \includegraphics[width=0.5\textwidth]{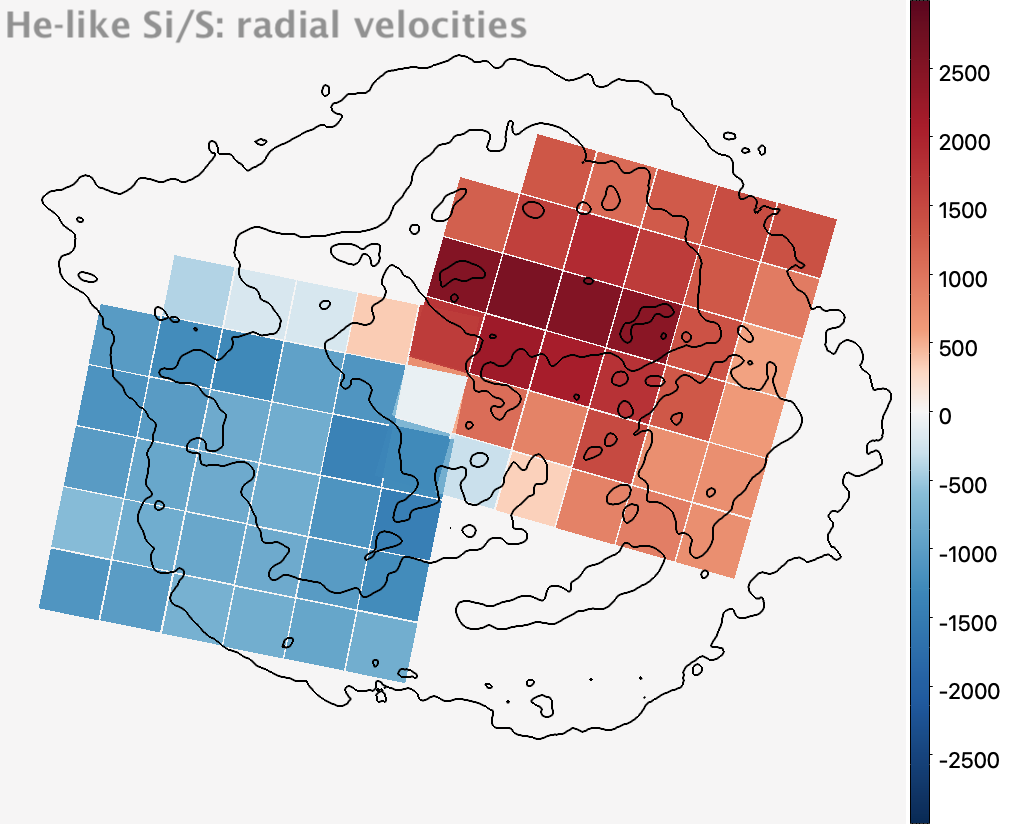}
    \includegraphics[width=0.5\textwidth]{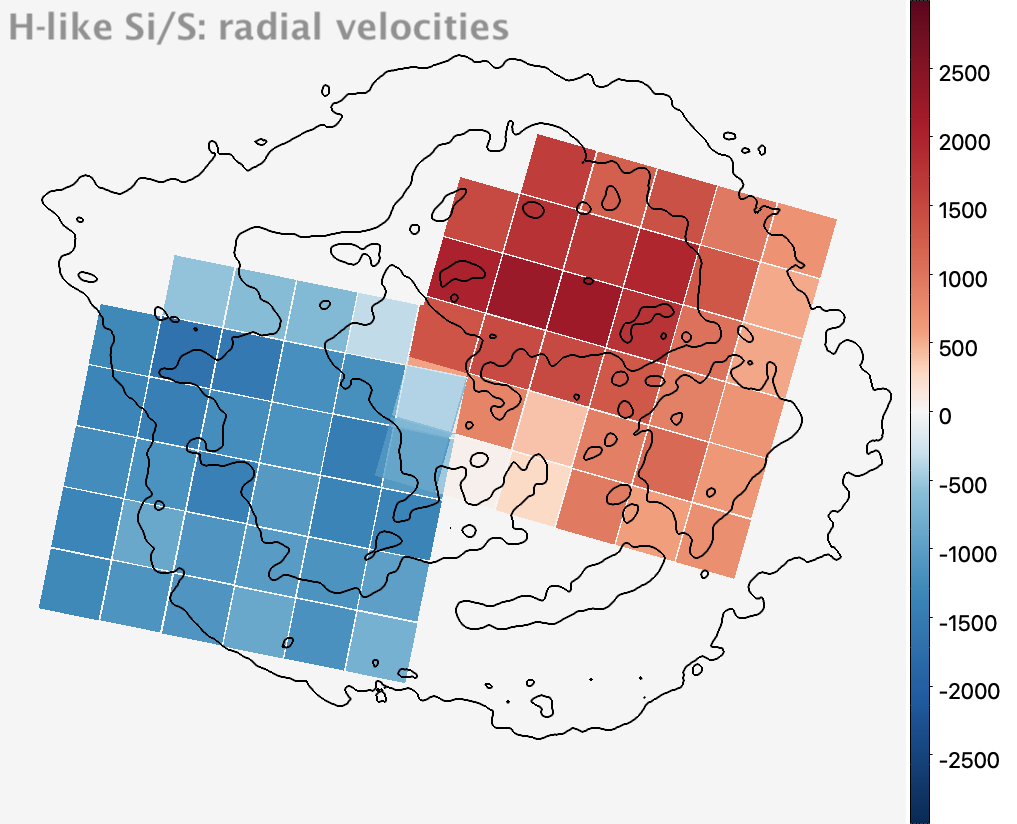}
  }
  \centerline{
    \includegraphics[width=0.5\textwidth]{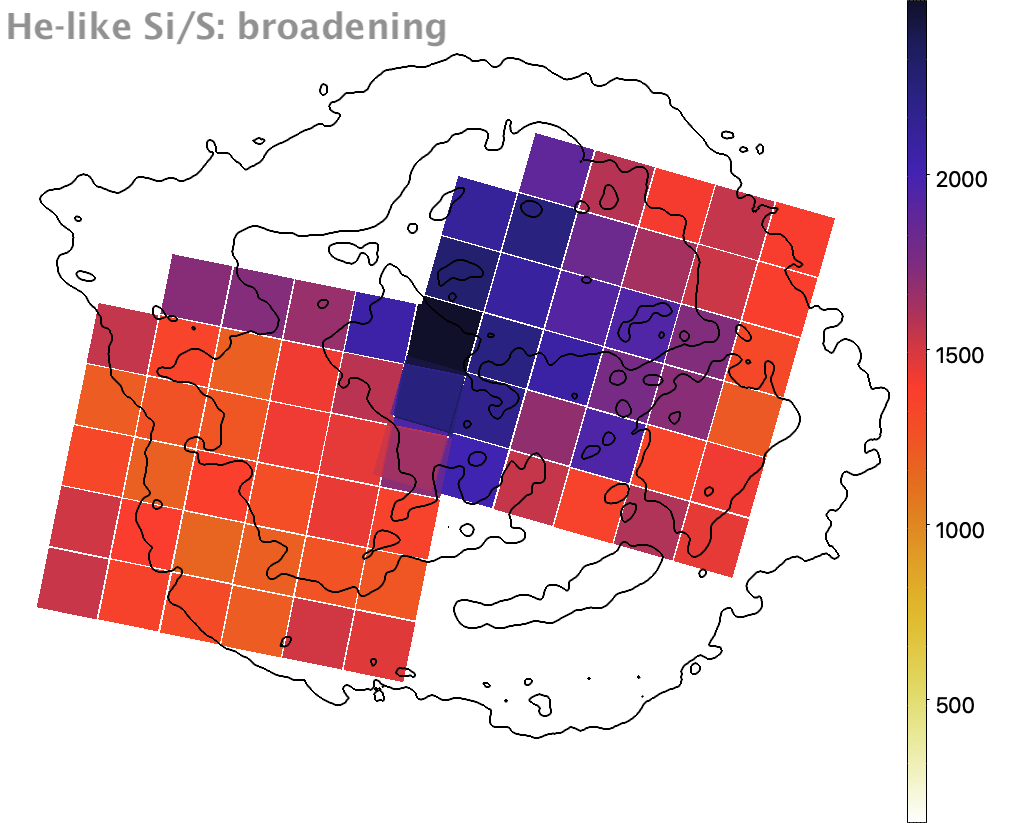}
    \includegraphics[width=0.5\textwidth]{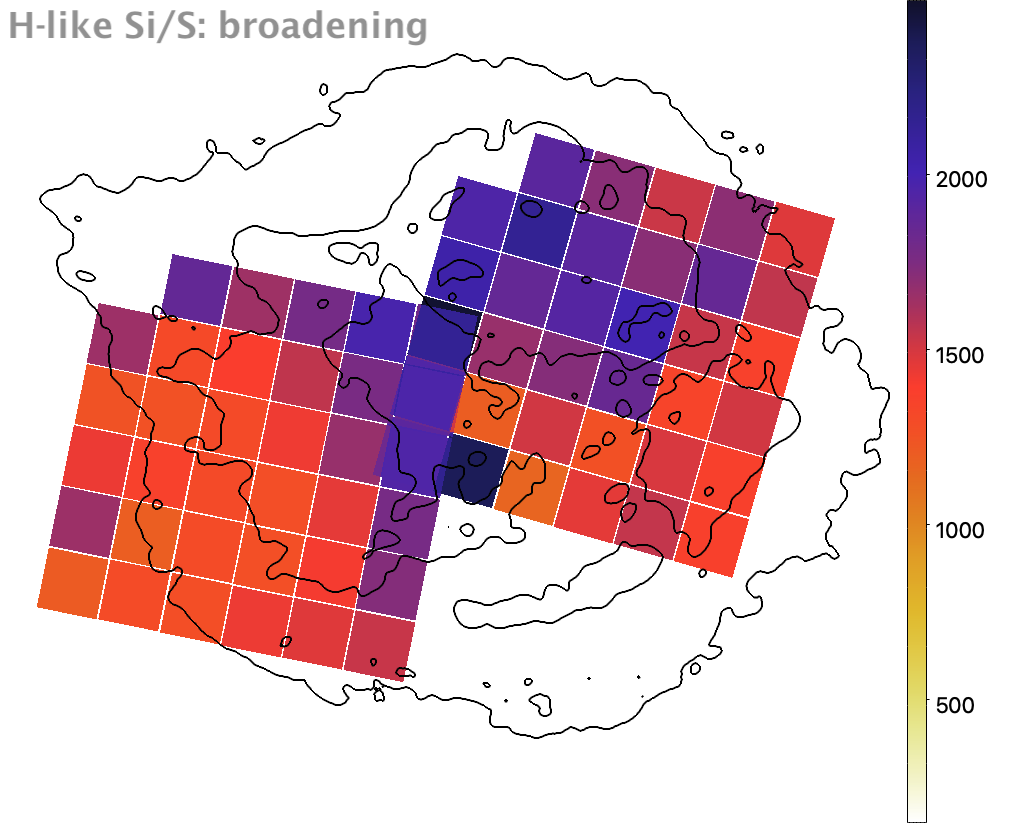}    
  }
  \centerline{
    \includegraphics[width=\columnwidth]{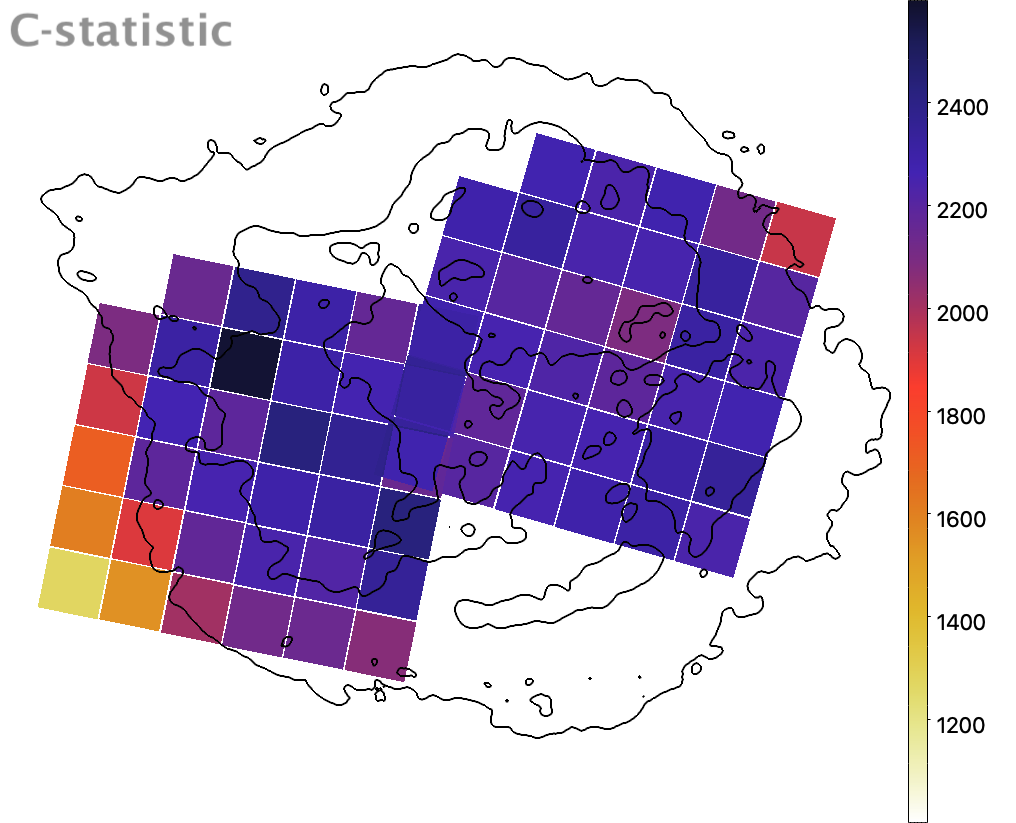}    
  }
  \caption{
    Maps of the results of fitting the spectra in the 2--3 keV line with Gaussian lines, seperating out the Doppler parameters for the He-like and H-like
    lines of silicon and sulfur.
    Top panel: Radial velocity maps of  He-like Si+S (left) and H-like Si+S (right).
    Middle panel: Doppler broadening ($\sigma_v$) maps of  He-like Si+S (left) and H-like Si+S (right).
    Bottom panel: C-statistic map. The number of degrees of freedom is 1980.
    The units for the color bars of the top four panels are \kms.
    The contours show the outline of Cas A as derived from the \chandra\ observation from 2019 (ObsID 19606), using the Si XIII emission from 1.75--1.97 keV.
{Alt text: Five maps showing the radial velocity parameters. Chandra X-ray contours are overlaid for orientation. Top panels: the radial velocities per pixel; left He-like lines, right H-like lines. Medium panels: line broadening measurements. Bottom panel: C-statistic.
}
    \label{fig:maps_Gaussians}
}
\end{figure*}

\begin{figure*}
  \centerline{
    \includegraphics[width=0.5\textwidth]{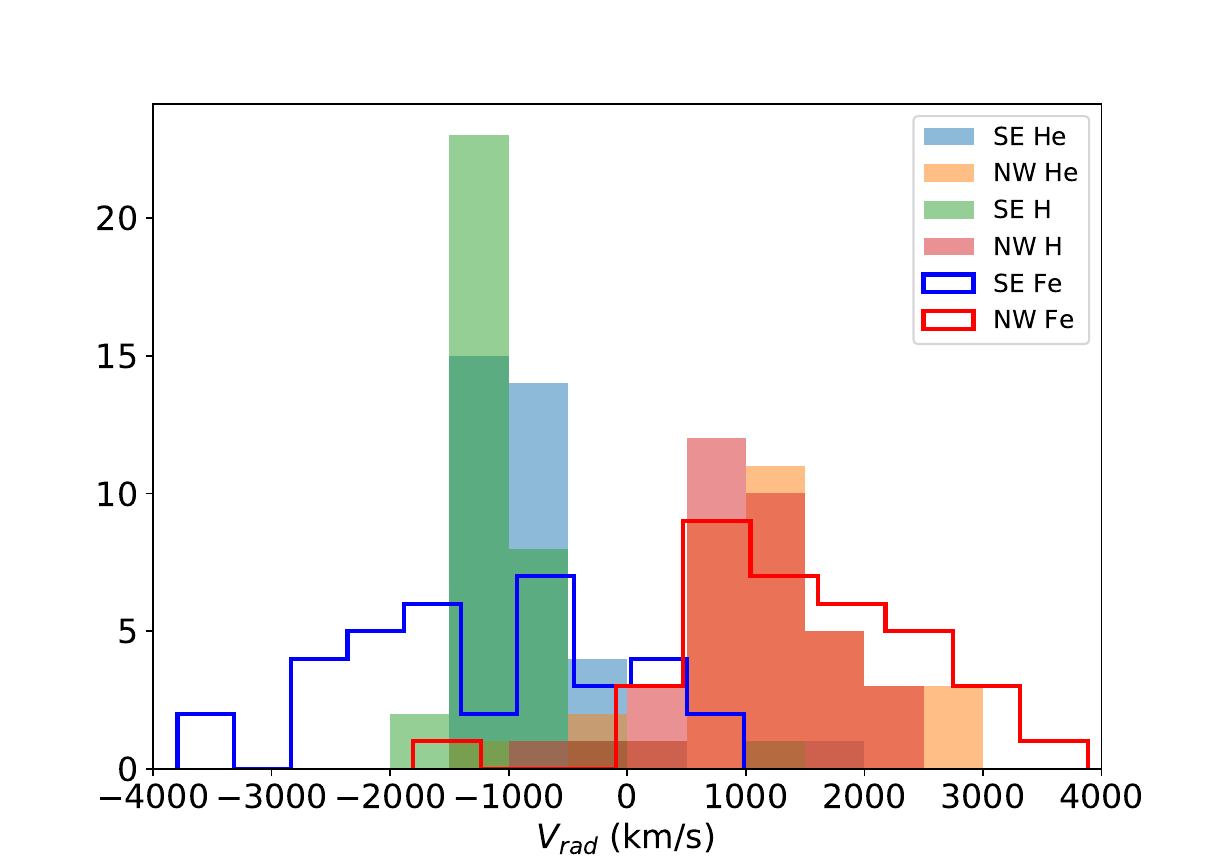}
        \includegraphics[width=0.5\textwidth]{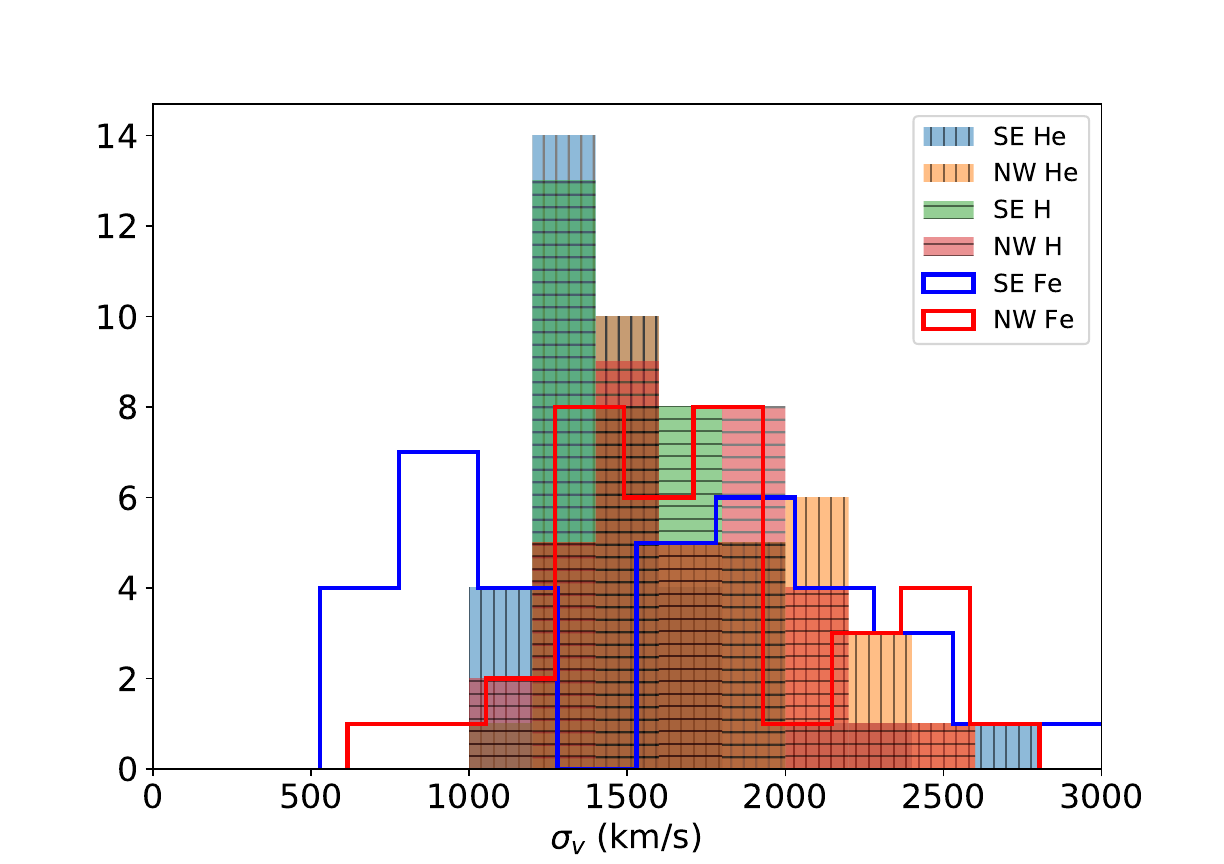}
    }
  \caption{
    Left: histogram of the radial velocity distributions for He-like lines and H-like lines of silicon and sulfur. 
    Right: idem, but now for the Doppler broadening, $\sigma_v$. In order to better separate the results from the SE and NW pointings,
    both are hatched differently.
    For comparison also the radial velocity and $\sigma_v$ 
    distribution based on fitting the Fe-K line complex are shown with a blue (SE) and red line (NW). See Bamba et al. 2025 for details.
    {Alt text: Left: Histogram of the radial velocity distributions for the SE and NW pointings, separated in color for He- and H-like lines.
    Right: histogram of the velocity broadening.
    } 
\label{fig:hist}
  }
\end{figure*}

\subsection{Automated fits with single Gaussian line components}\label{sec:automated}
 
With a spectral resolution of $\sim$5~eV \resolve\ can resolve individual lines, provided the Doppler broadening is not too large, and the lines are not too
narrowly packed. The \resolve\ spectra of Cas~A  indeed resolve most individual lines between 1.8 and  5 keV, dominated by He-like and H-like emission from
silicon, sulfur, argon and calcium.
However, the Fe-K complex consists of various charge states, and cannot be studied using individual lines.
For that reason in this paper we concentrate on the Si and S emission between 2-3 keV. 
A separate Doppler study of the the Fe-K complex
is reported in \citet{bamba25}, employing a full non-equilibrium ionization plasma model.
In the present study we rely on measuring simultaneously all lines Si and S lines using Gaussian line models, with a power-law spectrum to
heuristically model the underlying continuum. The spectral fitting was done with automated scripts generating models for the \xspec\ v12.12 \citep{arnaud96}
software package, employing the Poissonian maximum likelihood method, or C-statistic \citep{cash79} as a goodness of fit measure.
 Altogether the model consisted of 19 Gaussian components--modeled using the {\it zgauss} model in \xspec. 
 Note that the He$\alpha$ triplet emission of Si and S is not or barely   resolved, but as its broadening
 and radial velocity information is coupled to the resolved He$\beta,\gamma$ lines, we did retain the He triplet
 into our spectral fits.
 
The most prominent components for modeling the lines are the following:
Si Ly$\alpha$ (2.005 keV), 
Si He$\beta$ (2.183~keV),
Si He$\gamma$ (2.293~keV),
Si He$\delta$ (2.346~keV),
Si Ly$\beta$ (2.377 keV),
the S He$\alpha$ triplet (2.461 , 2.448 keV, 2.430 keV),  
Si Ly$\gamma$ (2.506 keV),
S Ly$\alpha$ (2.622~keV), and
S He$\beta$ (2.884~keV).
The Ly$\alpha$ lines and the intercombination lines of He$\alpha$ do in fact consist of two narrowly separated
components, which we ignored here, as even without line broadening \resolve\ is not able to separate them \citep[e.g.][]{gunasekera25}.
We also added two Gaussian components that likely correspond to He$\beta$ satellite lines,
from double excited Li-like ions at 2.134 keV \citep[Si, see e.g.][]{phillips06} 
and 2.829 keV (S).
These line transitions likely result
from inner shell ionization of Be-like ions or from dielectronic recombination of He-like ions. 
We found that the prominence of these line features varies considerably from region to region.

During the exploration of the spectra evidence was found that H-like and He-like lines have systematically different Doppler shifts.
The reason is most likely due to velocity differences in plasma shocked at different times, causing Doppler-shift
variations depending on ionization age, as reported in \citet{suzuki25}.
For that reason, the redshift ($z$) and line broadening ($\sigma$) parameters of all He-like lines were coupled together and similarly those for all H-like lines.

In Fig.~\ref{fig:spectra} two examples of automated fits are shown. In general, the spectra are well fitted with C-statistic values ranging from
1271 to 2581 for 1980 degrees of freedom. Fig.~\ref{fig:maps_Gaussians} shows maps of the measured radial velocities ($v_{\rm rad}$, top) and line broadening ($\sigma_v$, middle)
 for the He-like (left) and H-like ions (right). For overlapping SE/NW pixels, the averages are shown.
 The bottom panel shows the C-statistic values. In general, the C-statistic values scale with the statistical quality of the spectra,
 with a particular low value for the lower-left corner pixel in the SE, which is mostly based on events coming from photons scattered by the PSF. We iterate here  our earlier statement that the spectra from the SE corner of the SE pointing, and the NW corner of the NW pointing are dominated by photons scattered by the PSF from the bright shell.

 The velocity maps confirm the earlier results obtained with the Einstein FPCS and CCD detectors \citep{markert83,holt94,willingale02}: the SE region shows plasma that is blueshifted, whereas the NW region is overall redshifted.
 Despite a much better spectral resolution, the radial velocities do not show any new surprises, but we are now much more confident that the results are not influenced
by systematic uncertainties caused by the fact that we resolve individual lines rather than relying on centroiding of line features consisting of multiple lines.

The new capability is that we can measure the individual
 line broadening ($\sigma_v$), which could be caused by velocity variations along the line of sight, enhanced by angular blending due to the PSF,
 or can be due to thermal Doppler broadening; see \sect~\ref{sec:sigma_v} for further
 discussion.
 We find a $\sigma_v$ range of $\sim 1150~{\rm km\,s^{-1}}$ to  $\sim 2630~{\rm km\,s^{-1}}$ for both
 the H-like and He-like lines. 
 The highest $\sigma_v$ is found in the central region, where indeed we expect the largest
 broadening due to line of sight effects. Toward the edges, the intrinsic line broadening, i.e. thermal line broadening, should be more dominant. In that light it is interesting that toward the edge
 of the FoV in the SE pointing we still measure $\sigma_v\approx 1700~{\rm km\,s^{-1}}$ at the base of the NE jet (pixels 11 and 9).

Fig.~\ref{fig:hist} shows the histograms of the measured Doppler velocities and broadenings. The histogram of the radial velocities shows that for the SE region the range of velocities
is much more limited---$-1678$ to $79~{\rm km\,s^{-1}}$, or $\Delta v\approx 1750~{\rm km\,s^{-1}}$---than for the NW region: 
$-303$ to $2604~{\rm km\,s^{-1}}$, or $\Delta v\approx 2900~{\rm km\,s^{-1}}$. We indeed find a clear distinction in radial velocities for the He-like and H-like lines.
But in the SE region the H-like lines show the highest absolute radial velocities, whereas for the NW region the He-like lines show the highest absolute radial velocities.

For comparison we also show the Doppler measurements based on modeling the Fe-K emission from Cas A with \resolve\ \citep[solid lines][]{bamba25}. Compared to Si, S, the Fe K lines show a much
wider range in $v_{\rm rad}$ and $\sigma_v$. This was also reported by \citet{willingale02}. 
This likely hints that there are more iron protrusions in Cas A than just in the SE corner, where the protrusions are more conspicuous as they are projected beyond the Si-rich main shell.

	\subsection{Decomposing line shapes into two Gaussian components}\label{sec:double}

\begin{figure*}
 \centerline{
    \includegraphics[width=0.5\textwidth]{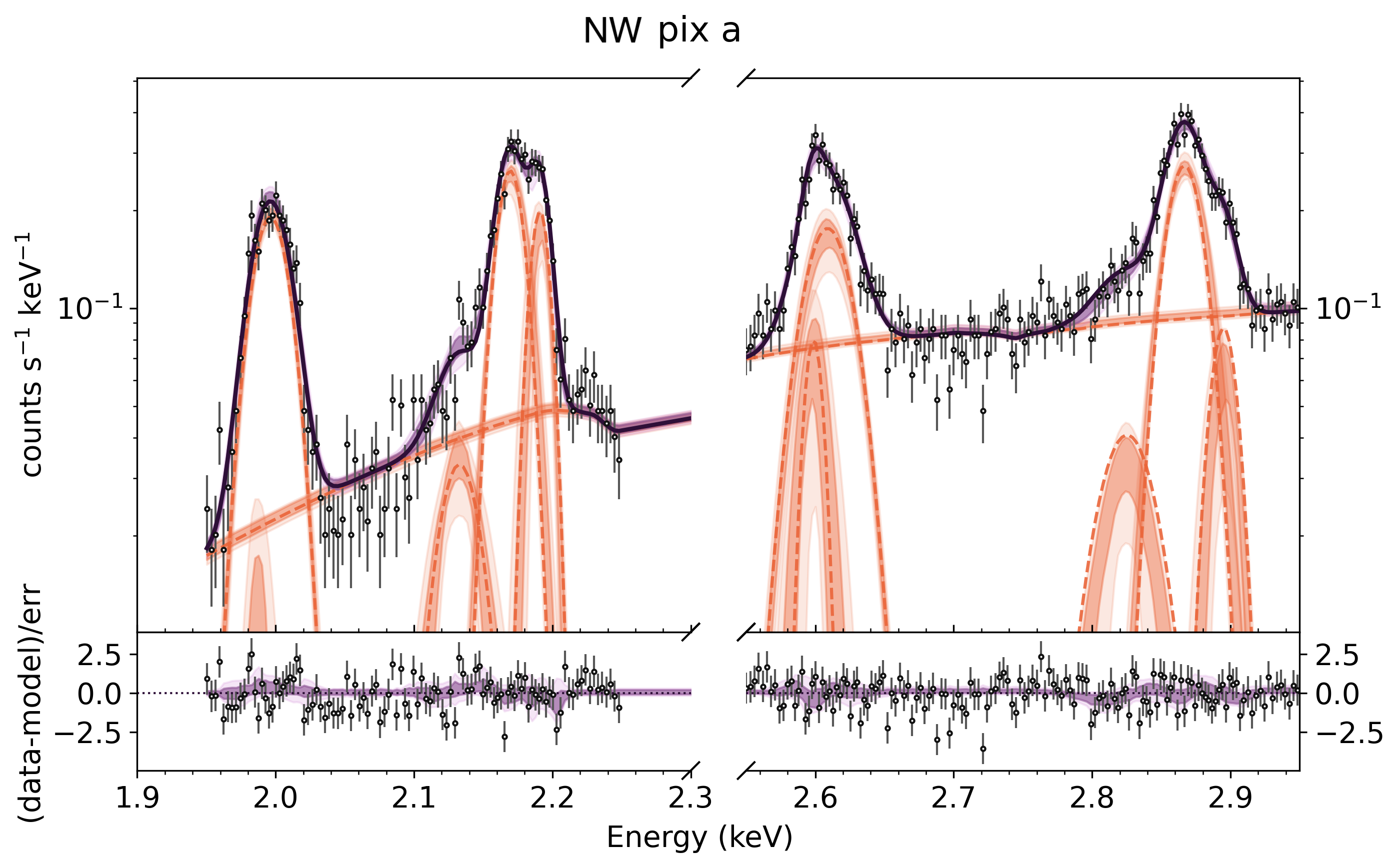}
   \includegraphics[width=0.5\textwidth]{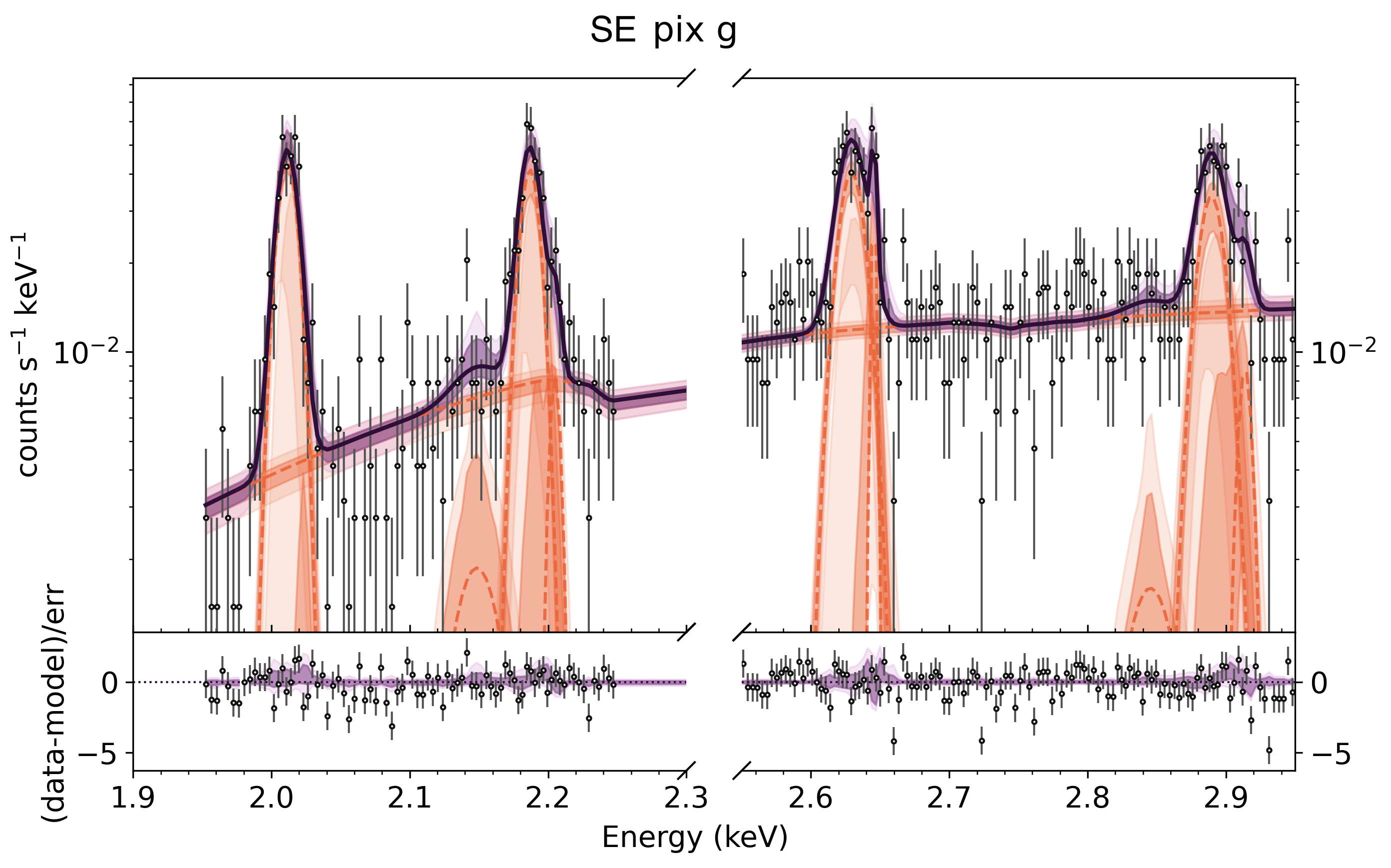}
    }
  \caption{
    Two examples of Bayesian spectral fitting of four bright Si and S lines with double Gaussians.
    From left to right:
    Si Ly$\alpha$, Si He$\beta$, S Ly$\alpha$ and S He$\beta$. In addition,
    weaker He$\beta$ satellite lines of Si and S were added, needed to
    fit the shoulder to the He$\beta$ lines.
    The orange lines show the individual components, with the thickness indicating the uncertainties in the individual line component. 
    Left: Spectrum of NW 2$\times$2 pixel {\em a}. For this spectrum fitting double Gaussians---instead of a single Gaussian---for each line substantially improved the fit, with $\Delta C=102$ for 8 additional degrees of freedom.
    Right: Spectrum SE 2$\times$2 pixel {\em g}.
    The improvement was marginal with $\Delta C=26$ for 8 additional degrees of freedom. 
    {Alt text: Two example spectra showing the Ly$\alpha$ and He$\beta$
    line emission, with the fitting results using two Gaussian components per line. Left: NW pixel a; right: SE pix g.}
    \label{fig:lineprof_ex}
  }
\end{figure*}

\begin{figure*}
  \centerline{
    \includegraphics[width=0.5\textwidth]{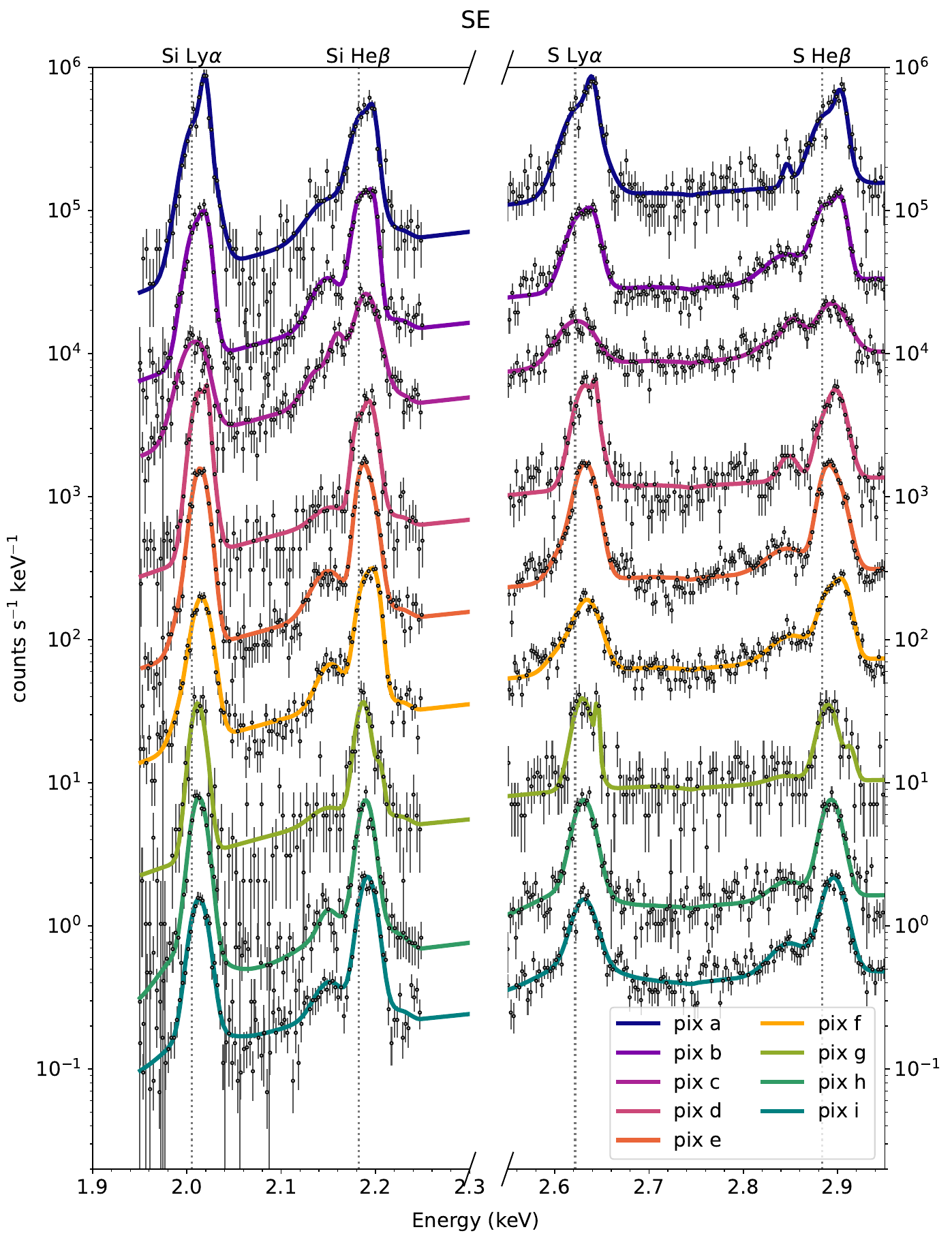}
    \includegraphics[width=0.5\textwidth]{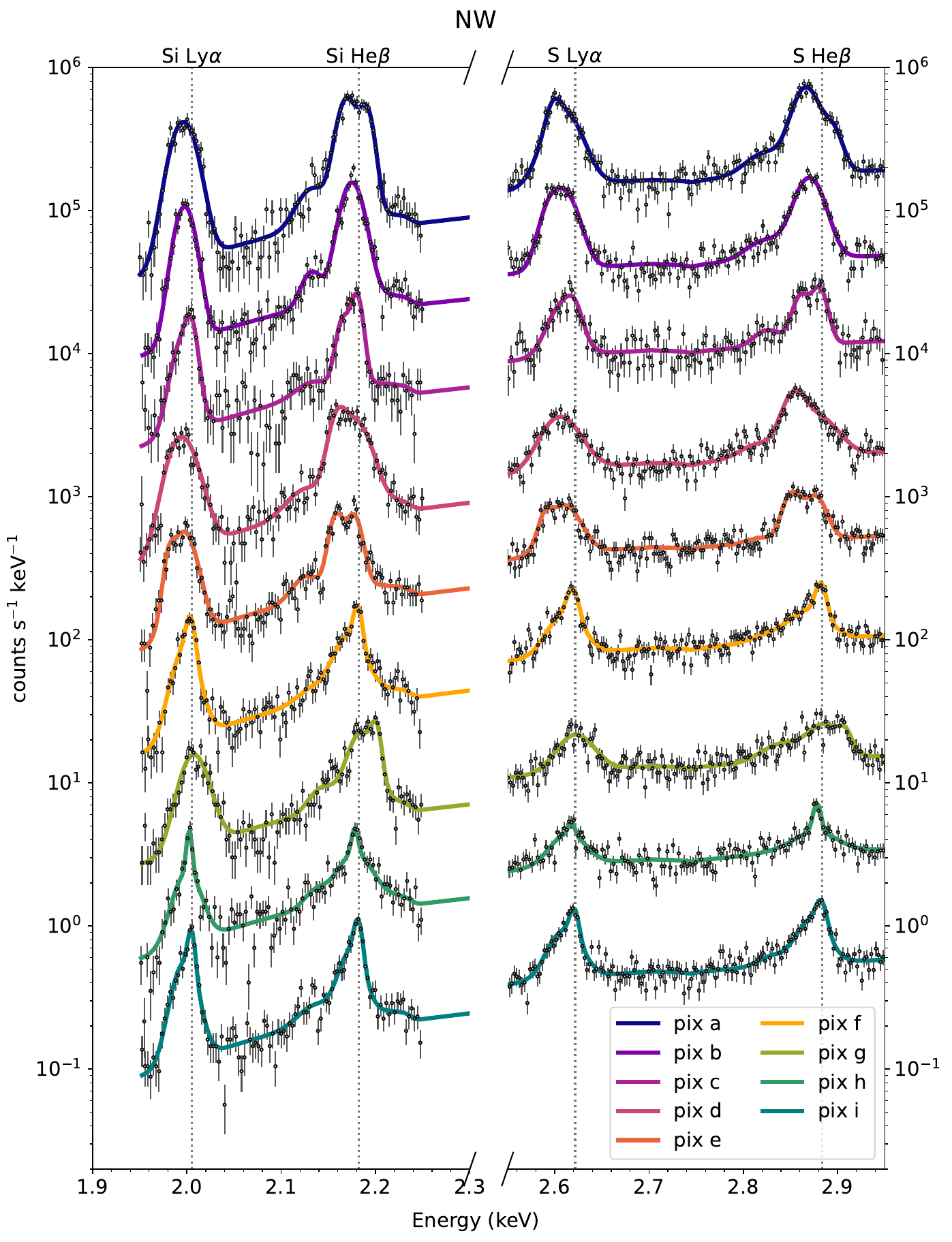}
    }
  \caption{
    Line profiles of Si Ly$\alpha$, Si He$\beta$, S Ly$\alpha$ and S He$\beta$ as fitted using two Gaussians. The fits are for the binned 2$\times$2 pixels (a,...,i). See Fig.~\ref{fig:lineprof_ex} for a detailed look of two examples.
    {Alt text: Line profiles of Si Ly$\alpha$, Si He$\beta$, S Ly$\alpha$ and S He$\beta$ for all pixels, and with best-fit two-component fits.} 
    \label{fig:lineprofs}
  }
\end{figure*}

\begin{figure*}
\centerline{
  \includegraphics[width=0.5\textwidth]{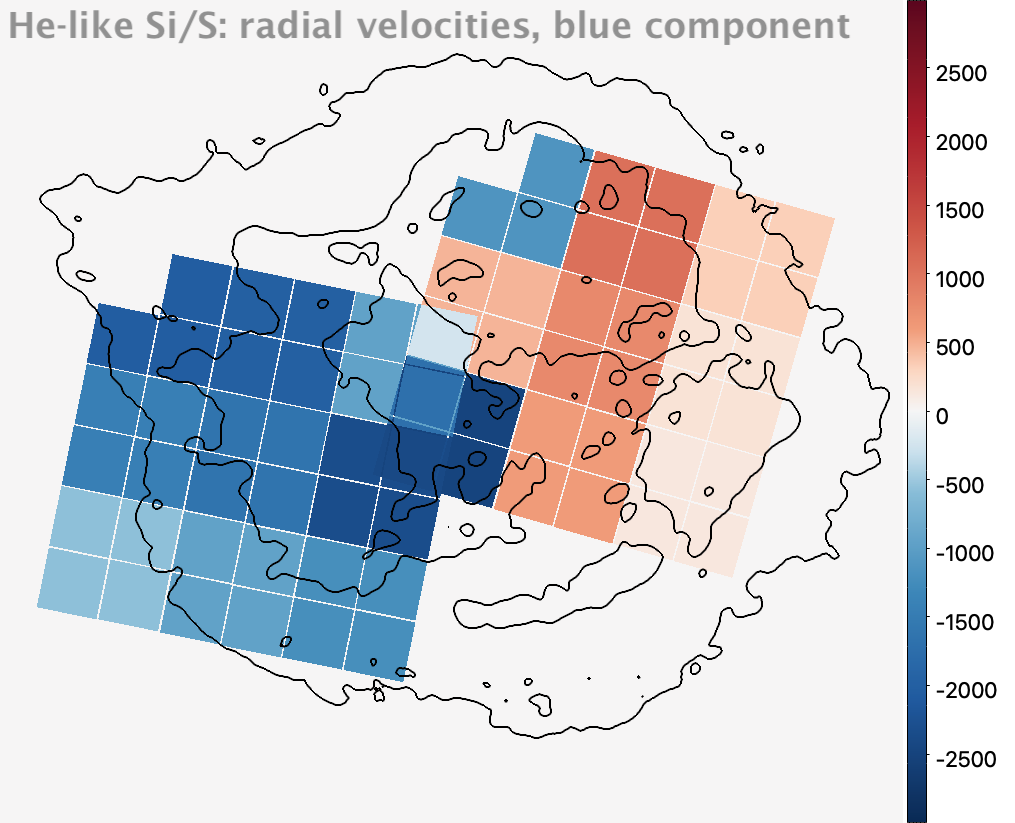}
  \includegraphics[width=0.5\textwidth]{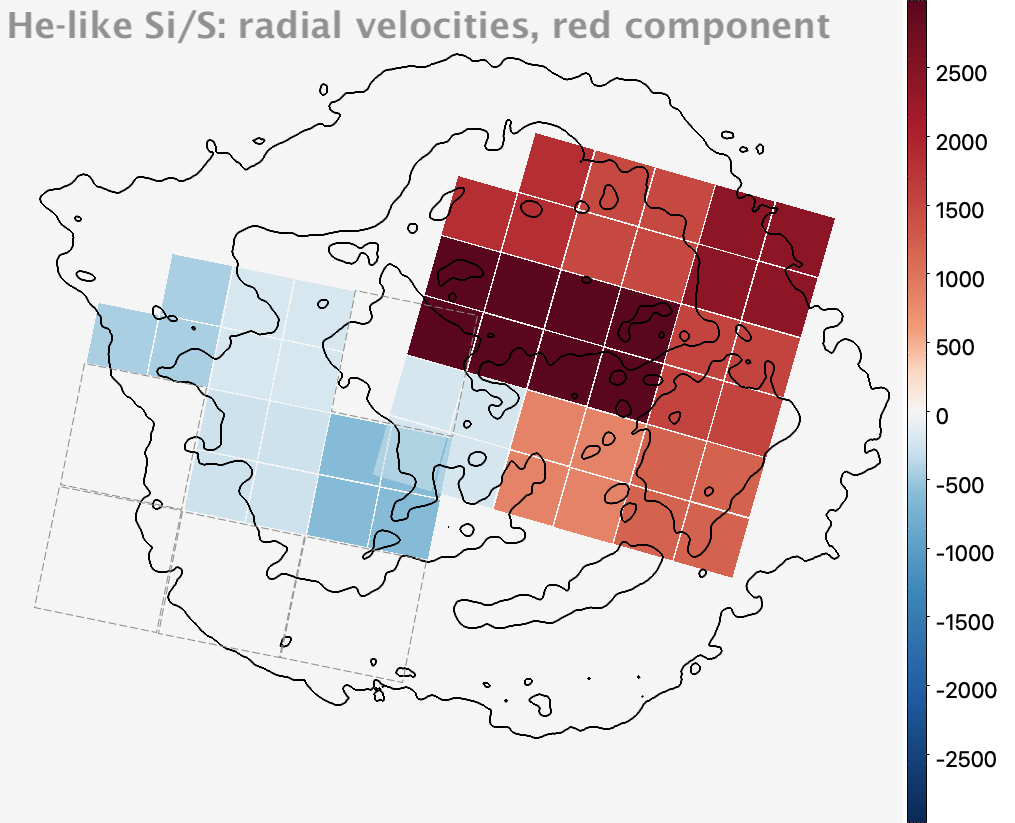}
}
\centerline{
  \includegraphics[width=0.5\textwidth]{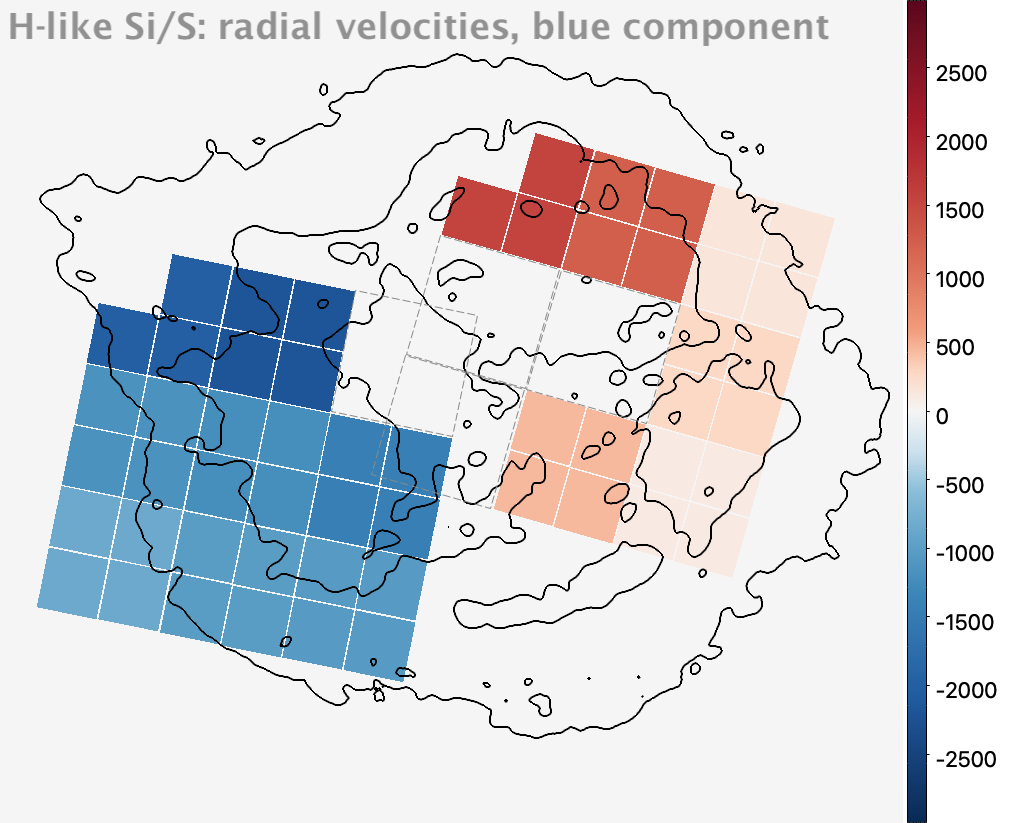}
  \includegraphics[width=0.5\textwidth]{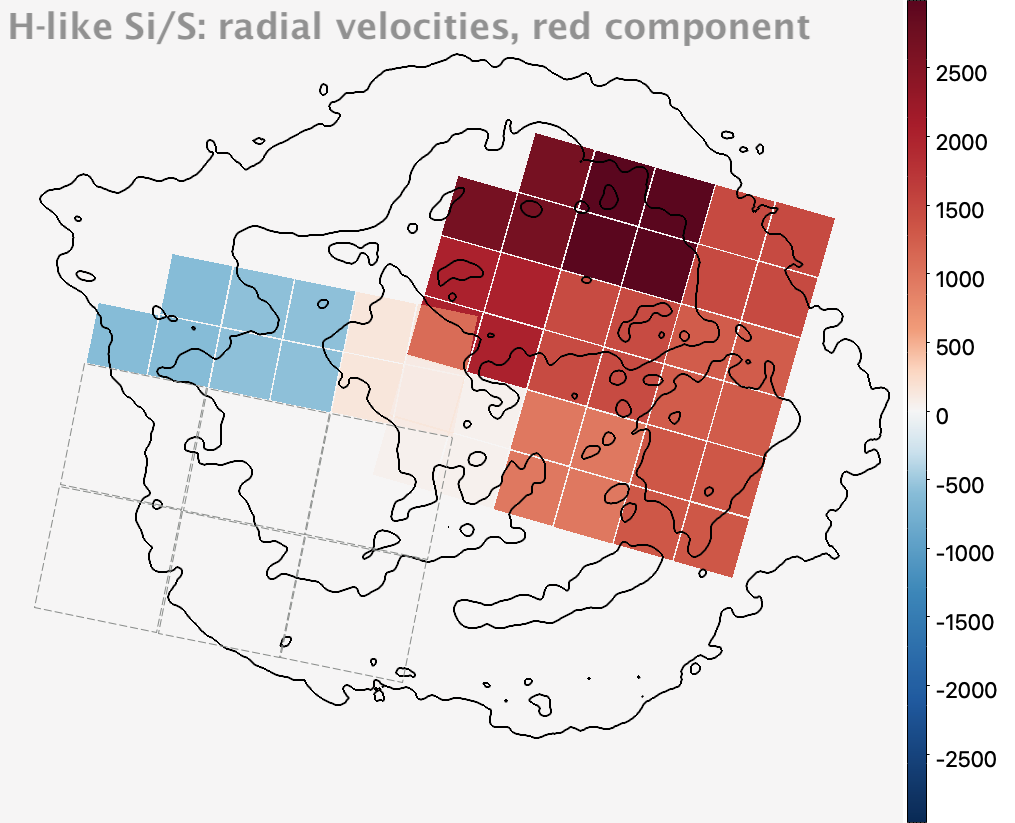}
}
\caption{
  Radial velocity ($v_{\rm rad}$) maps at 2$\times$2 pixel resolution of the two Gaussian components for the He$\beta$ lines (top) and Ly$\alpha$ (bottom) lines Si and S.
  On the left are the relatively blueshifted components and on the right the relatively redshifted components.
$v_{\rm rad}$ is only displayed for those components which significantly contribute---see text.
  The colorbars are identical to  those of the $v_{\rm rad}$ maps in Fig.~\ref{fig:maps_Gaussians},
    and the values correspond to \kms.
{Alt text: Four panels with radial velocity measurements using two-component
models. On the left: the relatively blueshifted components; on the right: the relatively  redshifted components. Top panels: He-like Si/S. Bottom panels: H-like Si/S.}
  \label{fig:vrad_maps_doublegauss}
}
\end{figure*}

\begin{figure*}
\centerline{
  \includegraphics[width=0.5\textwidth]{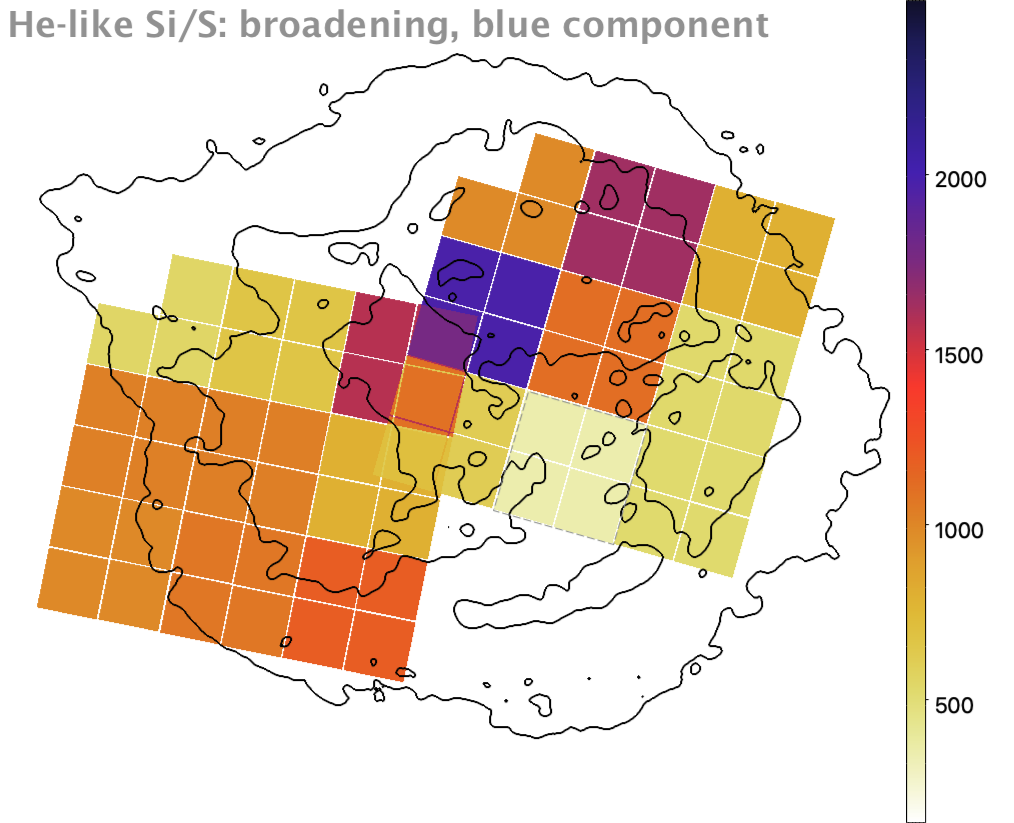}
  \includegraphics[width=0.5\textwidth]{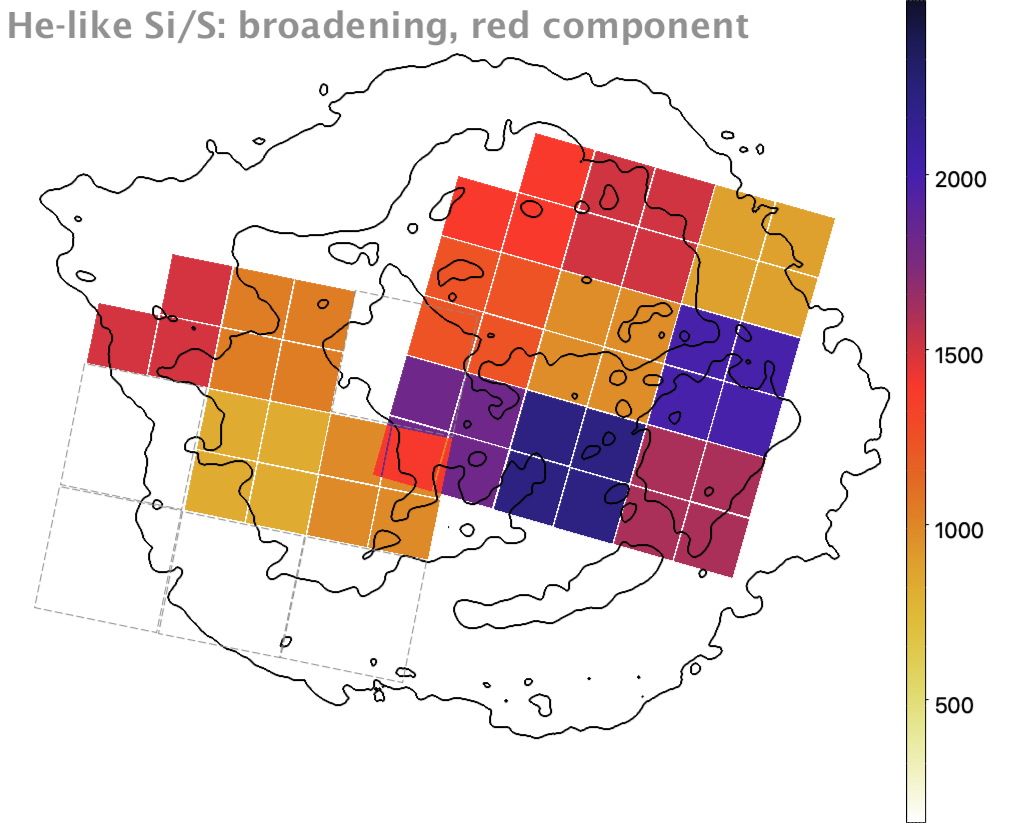}
}
\centerline{
  \includegraphics[width=0.5\textwidth]{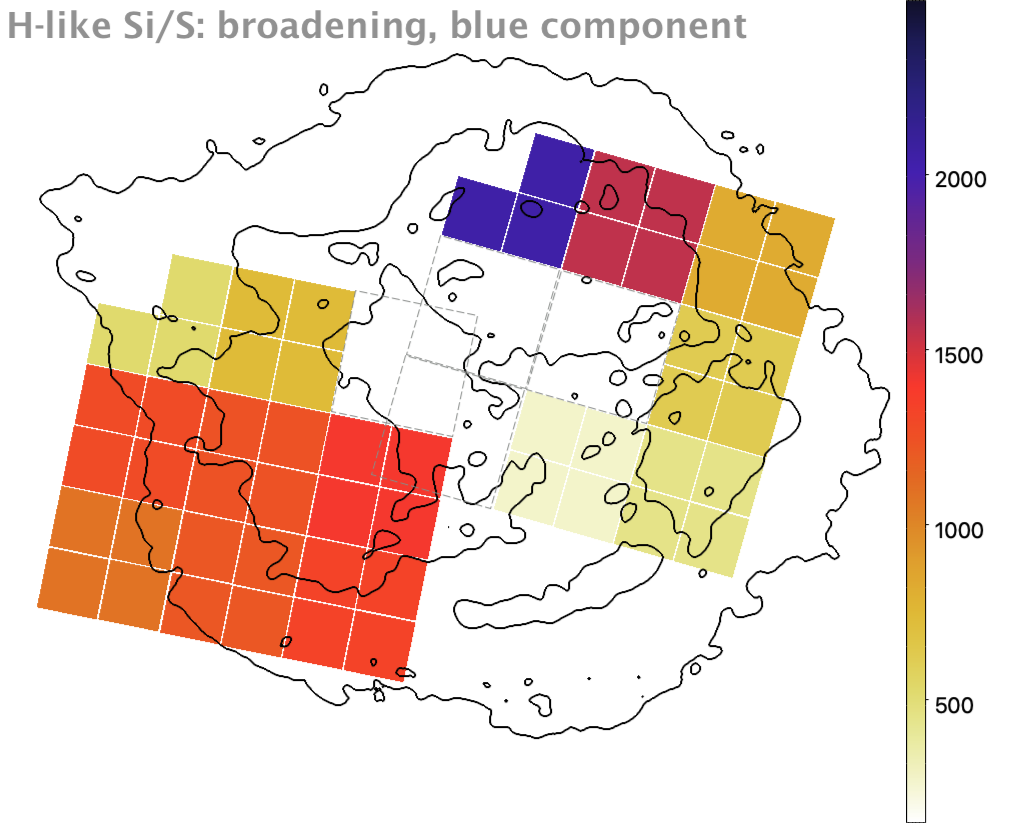}
  \includegraphics[width=0.5\textwidth]{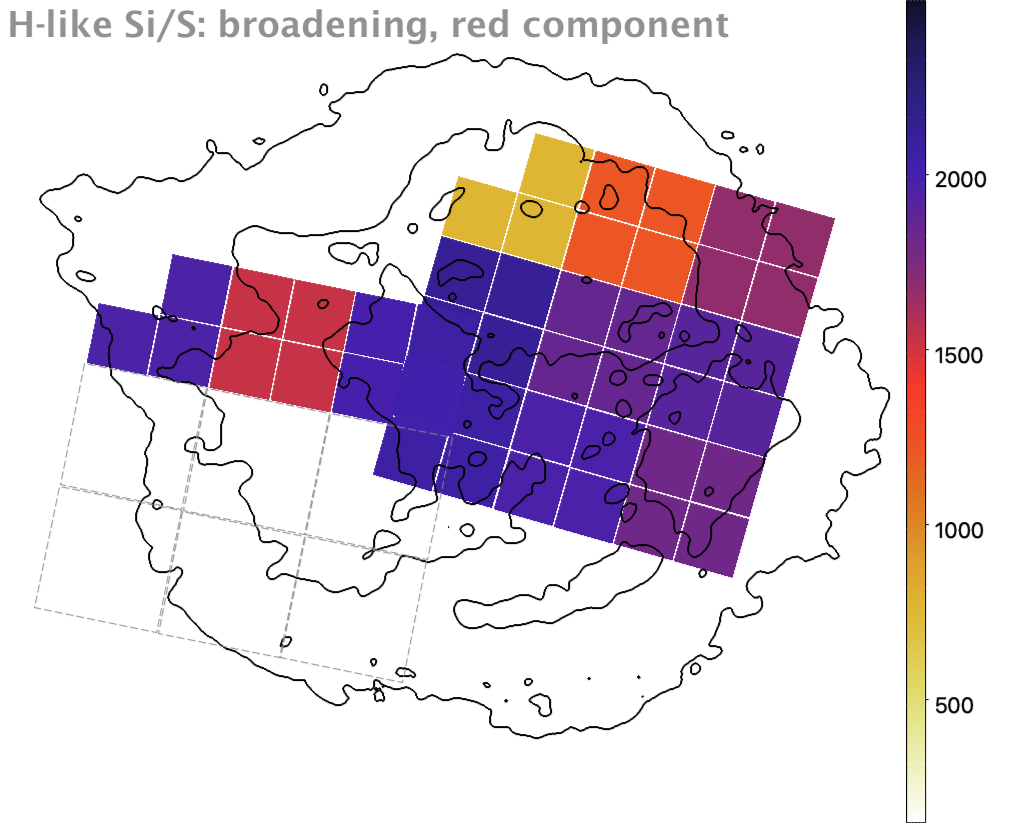}
}
\centerline{
  \includegraphics[width=0.5\textwidth]{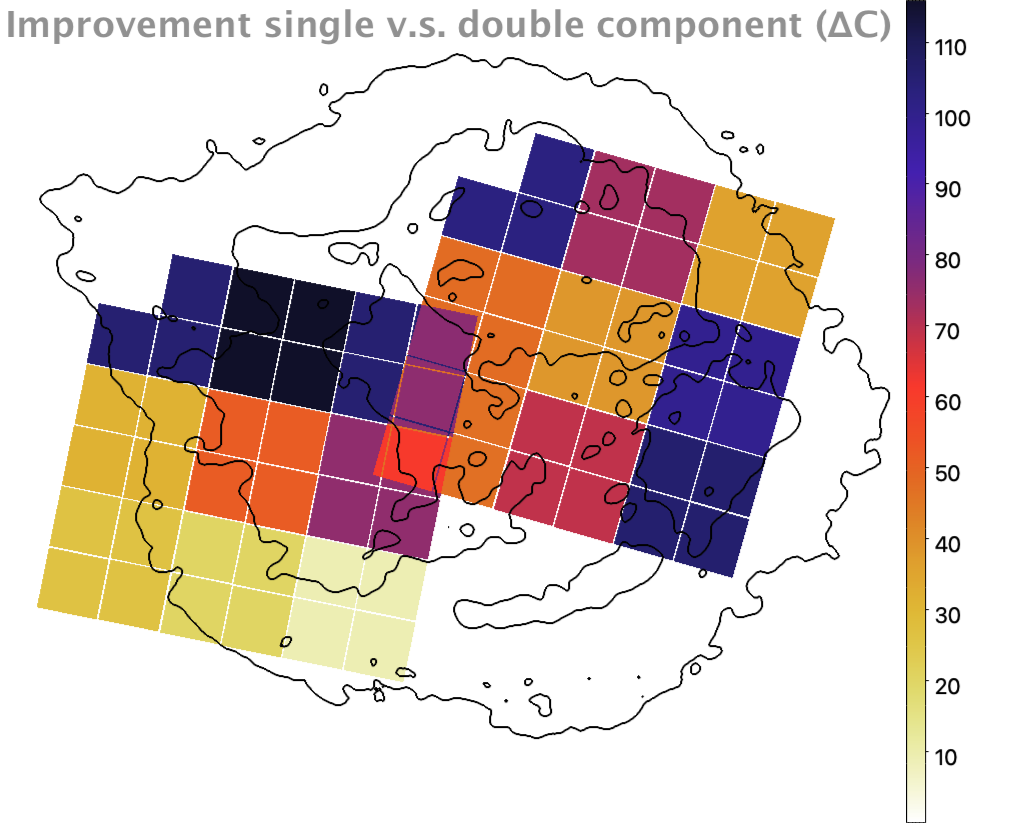}
  \includegraphics[width=0.5\textwidth]{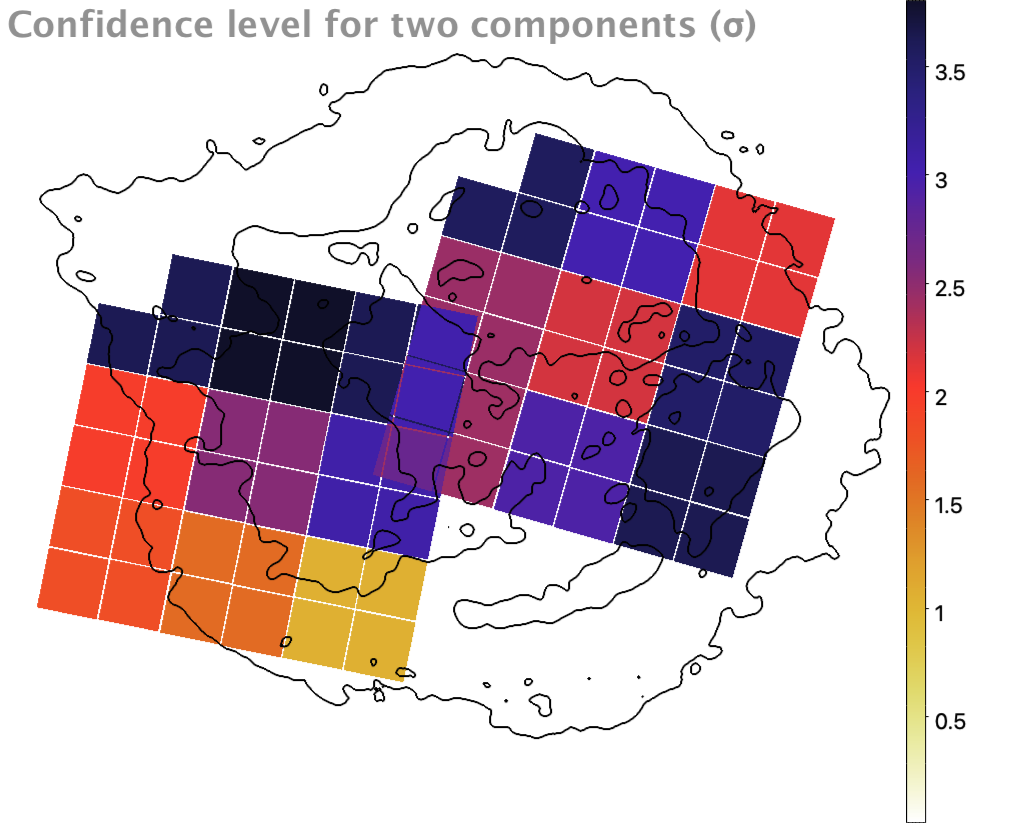}
}
\caption{
Top four panels:
  Doppler broadening ($\sigma_v$) maps at 2$\times$2 pixel resolution of the two Gaussian components for the He$\beta$ lines (top) and Ly$\alpha$ (central panels) lines Si and S.
  On the left are the relatively blueshifted components and on the right the relatively redshifted components.
$\sigma_v$ is only displayed for those components significantly contribute---see text.
  The colorbars are identical to  those of the $\sigma_v$ maps in Fig.~\ref{fig:maps_Gaussians},
  and the values correspond to \kms.
Bottom row, left: 
Maps of $\Delta C$-statistic, the statistical improvement obtained by fitting with two Gaussian line components compared to a single Gaussian component.
Bottom row, right: Improvement in terms of confidence level in $\sigma$.
{Alt text: Six panels with radial broadening measurements using two-component
models (top four panels) and two bottom panels showing the statistic improvements (left) and confidence levels (right). Top four panels: On the left: the relatively blueshifted components; on the right: the relatively redshifted components. Top two panels: He-like Si/S. Middle panels: H-like Si/S.}
  \label{fig:sigmav_maps_doublegauss}
}
\end{figure*}

The Doppler measurements reported in section~\ref{sec:automated} reveal with much more confidence and reliability the SE/NW asymmetry in X-ray Doppler velocities of Cas A, as well as providing
measurements of the line broadening. With X-ray line centroiding measurements one could still wonder if  variations in the He$\alpha$ line triplet ratios, or mixes of blueshifts and redshifts
left the impression of an overall blueshift/redshfift in Cas A, rather than a real kinematic difference.
However, fits with a single Gaussian component could still leave room for blends of truly
 redshifted and truly blueshifted components,
corresponding for the SE to emission from the near-side of the shell and a less prominent redshifted component from the far-side of the shell, and vice versa for the NW.

In order to investigate this, we analyzed in more spectral detail the Si and S Ly$\alpha$ and He$\beta$ lines, as these are single lines (as fine structure lines are not resolved at \xrism\ resolution, \cite{gunasekera25}) rather than complexes (like He$\alpha$).
In order to measure whether the single lines can be decomposed into a relatively blueshifted and redshifted component, we used two Gaussian components per line, again assuming similar Doppler velocity and broadening parameters for the He-like and H-like emission.

These two-component Gaussian models should be regarded as a tool to approximate radial velocity gradients and/or two or more distinct radial velocity components. As detailed infrared and optical radial velocity maps show \citep[e.g.,][]{delaney10,milisavljevic13,alarie14}, the three-dimensional distribution of ejecta knots is complex, which is likely to be also true for the more tenuous X-ray emitting ejecta.
A priori one would expect for each pixel to see both a redshifted
and blueshifted components from the far-side and the near-side of the shell.

We heuristically model the continuum using a power-law spectrum along with two Gaussians to capture Si and S He$\beta$ satellite lines, which can be present toward the low-energy tail of the He$\beta$ lines.

We found that an automated fitting procedure using similar scripts as used for section~\ref{sec:automated} does not lead to reliable results with two Gaussians. Instead we turned to the Bayesian X-ray Analysis 
scheme \citep[BXA v4.1.1][]{buchner14} with the {\em Ultranest (v4.3.2)} parameter sampling scheme \citep{buchner21} for reliable results. As BXA samples the entire parameter space, the procedure
is slow, given that the high resolution \resolve\ spectra require very large response files ($\sim 0.8$~GB). Moreover, for spectra from individual \resolve\ pixels, some regions of the remnant lack sufficient photon counts to reliably decompose the lines into two Gaussians. In order to overcome these problems, we applied BXA to the two times 9 ``super pixels" {\it a}--{\it i} for each pointing,
and we used the \texttt{SPEX v3.08.01} spectral fitting package \citep{kaastra96} with a spectral file format allowing to optimize the spectral binning \citep{kaastra16} and reducing the size of the spectral response files. For more details on Bayesian based spectral fitting using \texttt{SPEX} we refer to \cite{agarwal25}.
The larger region size also reduces the amount of spatial mixing caused by the broad \xrism\ PSF \citep{plucinsky25}.

The procedure was to first fit a single Gaussian component with BXA and then use the measured redshift $z_0$ and the corresponding error $\sigma_{z_0}$ to divide the line into two Gaussian components, 
constraining them such that one is a relatively blueshifted component and the other is a relatively redshifted component. This is achieved by setting a uniform prior on redshift in the range $z_0-10^{-2}$ to $z_0 + 2\sigma_{z_0}$ and $z_0 - 2\sigma_{z_0}$ to $z_0+10^{-2}$ for the relatively blueshifted and redshifted components respectively. As the redshift of these Gaussian components is defined relatively, it is possible that a component labeled relatively 'blueshifted' has a positive velocity (i.e. redshifted) and vice a versa. The broadening for both components has a uniform prior in between 100 \kms\ and 4000 \kms, allowing to explore scenarios like a blend of a narrow and a broad Gaussian as best-fit models. The normalizations of all components were left free. This also allowed for very low normalizations, which essentially meant that the single-Gaussian-component case
was included as a possible outcome. Due to the sampling of the parameter space, one can estimate the errors on the parameters using the posterior parameter distributions. 

The results of the BXA sampling were further curated manually, in that we selected the single Gaussian model
in case the improvements in statistical goodness of fits indicated no significant improvement ($\Delta C \leq 32$, i.e. a 2$\sigma$ improvement),
or in case a line normalization was very low compared to the dominant component (ratio of normalization less than 0.2). In one case, SE pixel {\it c}, the two component fit was significantly
better, but inspection of the fit revealed that the redshifted component of He$\beta$ lines was in fact fitting
the Si and S He$\beta$ satellite lines. The reason is that the He$\beta$ satellite lines are relatively bright for this pixel,
perhaps because of a low-ionization component. We therefore did not include this component in our
selection. 

Another case worth mentioning is the NW pixel {\it b}, for which
the Si and S He$\beta$ lines are best fitted with two components, with one component mostly
fitting the Si emission, and the other component mostly fitting the S emission. So, here the Si line emission is captured by the relatively blueshifted component and the S line by the redshifted component.  We included the two components in our selection.

In Fig.~\ref{fig:lineprof_ex} we provide two Bayesian model fits with the uncertainty bands for each component representing 95.4 percent posterior distribution range. 
\footnote{We will provide all data and figures in a repository on \url{https://zenodo.org} under \url{https://doi.org/10.5281/zenodo.14734013}.} 
On the left a fit for which two Gaussian components per line provide a significantly better fit than a single Gaussian component, and another one for
which a double Gaussian component does not provide a significant improvement.
Fig.~\ref{fig:lineprofs} provides an overview of all 18 BXA fits.
In order to facilitate a better comparison of the line shapes of the four lines for a given superpixel, we present an alternative version of the line profiles as function of radial velocity in the appendix, see Fig.~\ref{fig:lineprofiles_kms}.

\begin{figure*}
  \centerline{
    \includegraphics[width=0.5\textwidth]{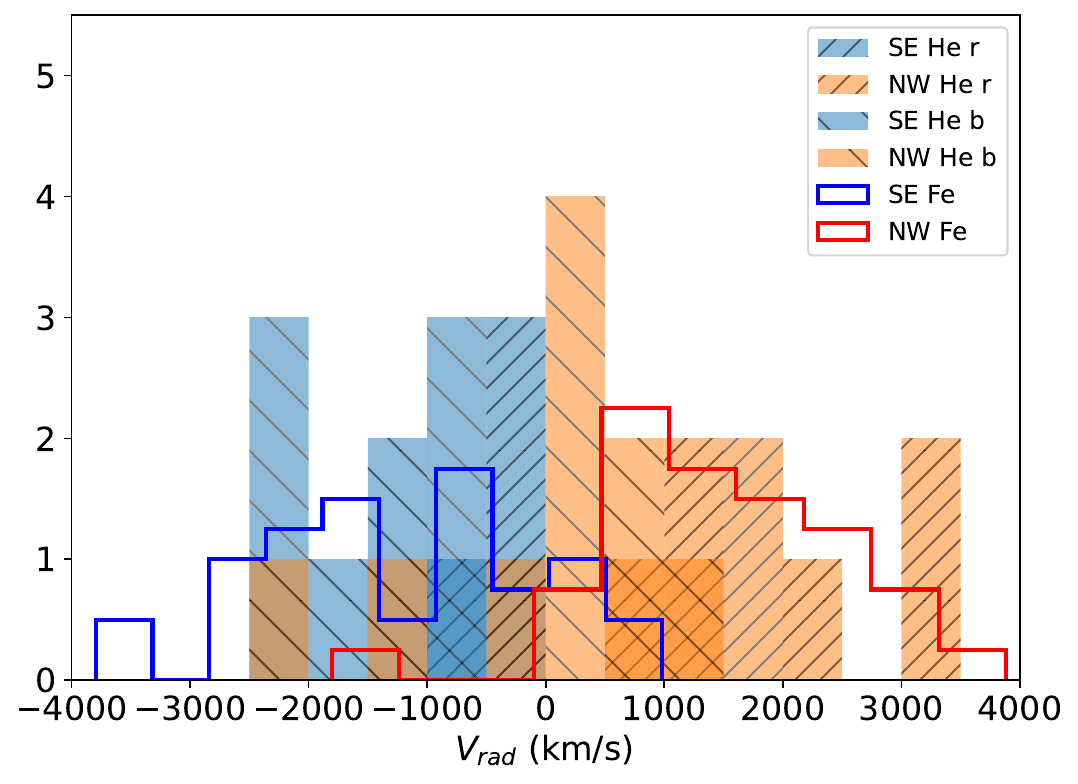}
    \includegraphics[width=0.5\textwidth]{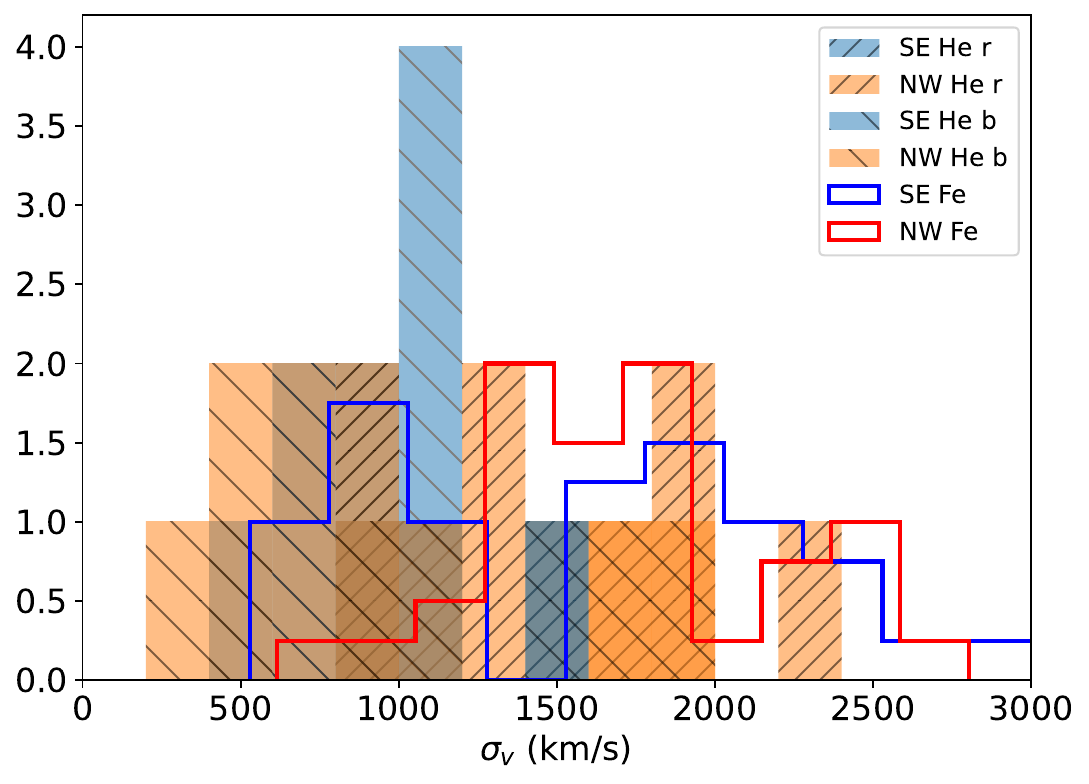}
    }
    \centerline{
    \includegraphics[width=0.5\textwidth]{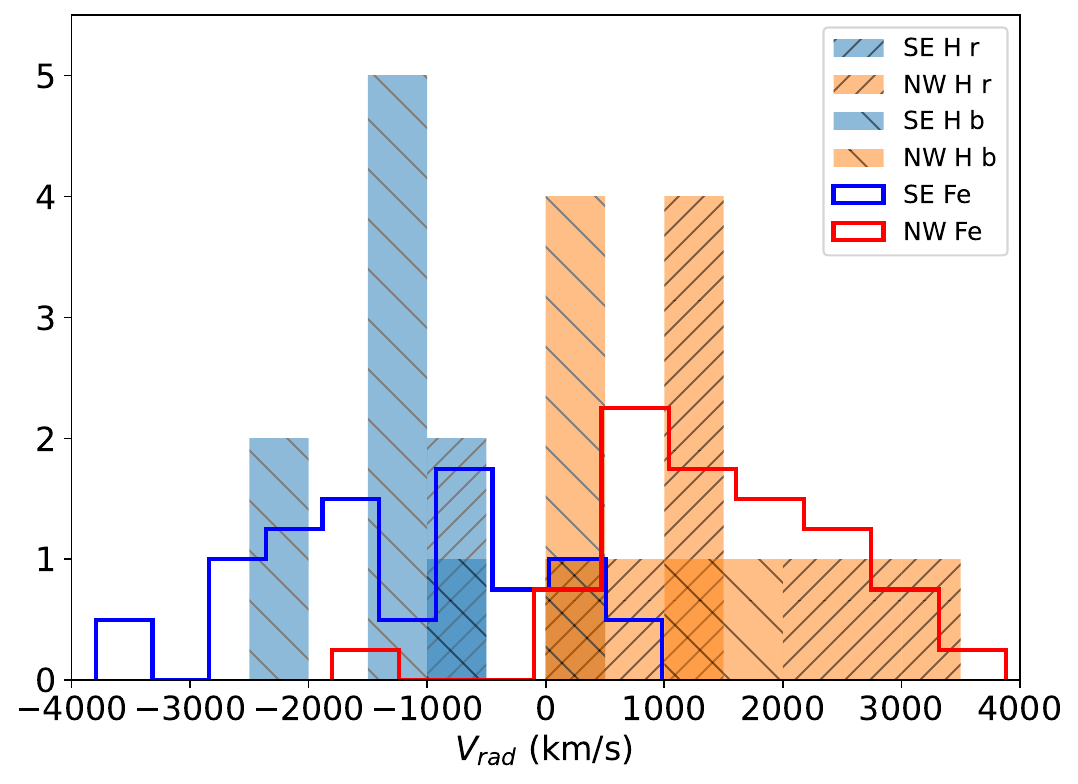}
    \includegraphics[width=0.5\textwidth]{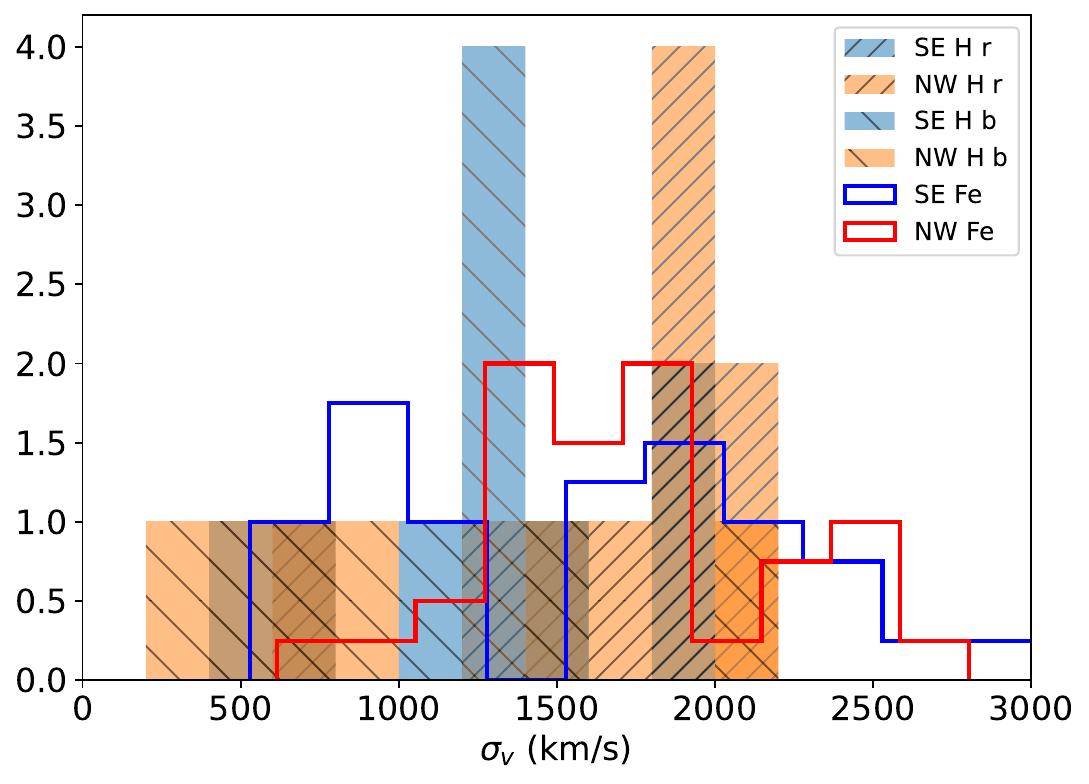}
    }
  \caption{
    Histograms of the radial velocities (left panels) and velocity broadeninng (right panels)
    for fits with two Gaussian components. Top panels: for He-like ions. Bottom panels:
    for H-like ions.
    The color coding is similar to Fig.~\ref{fig:hist} with  light blue indicating the SE region and orange indicating the NW region. The lowest velocity component (labeled rc) is  indicated by upward hatching and the highest velocity
    component (bc) by downward hatching.
    For comparison also the Fe-K  $v_{\rm rad}$ distributions (scaled) are shown \citep{bamba25}, which
    are based on fitting an NEI model, using a single velocity component.
{Alt text:
Four panels with histograms. On the left are radial velocity histograms, with on the top He-like ions and at the bottom H-like ions. 
On the right: similar histograms but now for the line broadening.
}
    \label{fig:vrad_hist_2gauss_separated}
  }
\end{figure*}

Fig.~\ref{fig:vrad_maps_doublegauss} and Fig.~\ref{fig:sigmav_maps_doublegauss} show the resulting maps of the radial velocities and line broadening, $\sigma_v$, for all line components.
Note that non-significant components are not shown, and are displayed as white pixels.
The radial velocity maps show that the dichotomy between the SE (mostly blueshifted) and the NW (mostly redshifted) in fact exists for both the $z_{\rm r}$ and $z_{\rm b}$ components.
For example, for the He-like lines, in the SE region, all significant fits show that both the redshifted and blueshifted component of the lines are blueshifted.
For the NW region, we see something similar: both the redshifted and blueshifted component are overall redshifted. An exception here is the NW pixel {\it a}, for which the blueshifted component has indeed a velocity of
$1150~{\rm km\,s^{-1}}$ toward us. 
However, this is only true for the He-like lines.
Contrary to expectations, in general the two-component Gaussian components appear to trace plasma components that are either both at the near-side (SE pointing) or both at far-side (NW) of the SNR.

As to be expected, the line broadening parameters are reduced with respect to the single Gaussian case, ranging now from $\sigma_v\approx 300$~\kms~ (Ly$\alpha$ blueshifted component, NW pixel {\it h}) to  $\sigma_v\approx 2200$~\kms~ (Ly$\alpha$ redshifted component, NW pixel {\it d}).
Overviews of the radial velocity and broadening distributions are given in the histograms in Fig.~\ref{fig:vrad_hist_2gauss_separated}.

As noted before, caution must be taken to overinterpret the maps near the edges, where most of the detected photons originate from outside the corresponding sky region due the broad PSF of \xrism. 
One may suspect that the relatively large radial velocity for the ``red" component in pixel {\it c} in NW is due to PSF scattering. However, for this pixel the two immediately neighboring it---{\it b} and {\it f}, accounting for $\sim 50\%$ of the photons--- display lower radial velocities. For pixel {\it e} an even higher radial velocity of the ``red" component is measured, but the contamination from this region to pixel {\it c} is relatively small, $\sim 10\%$. Although pixel {\it c} covers the edge of the shell,  there may still be some line of sight motion in this region, if the plasma motion is directed away from us with an angle of $\sim 20^\circ$---this would result in  radial velocities of $\sim 1500$~\kms, if the space velocity is $\sim 4500$~\kms.

Near the center of the SNR---pixels {\it c} of the SE pointing and pixel {\it g} of the NW pointing---the results should be interpreted with caution: this region has a much lower brightness than the shell, and in fact for these pixels a large fraction of the detected photons originate from the sky region corresponding to pixel {\it b} of the SE pointing.

\section{Discussion}
 
\begin{figure}
  \centerline{
   \includegraphics[trim=0 0 0 0 0,clip=true,width=\columnwidth]{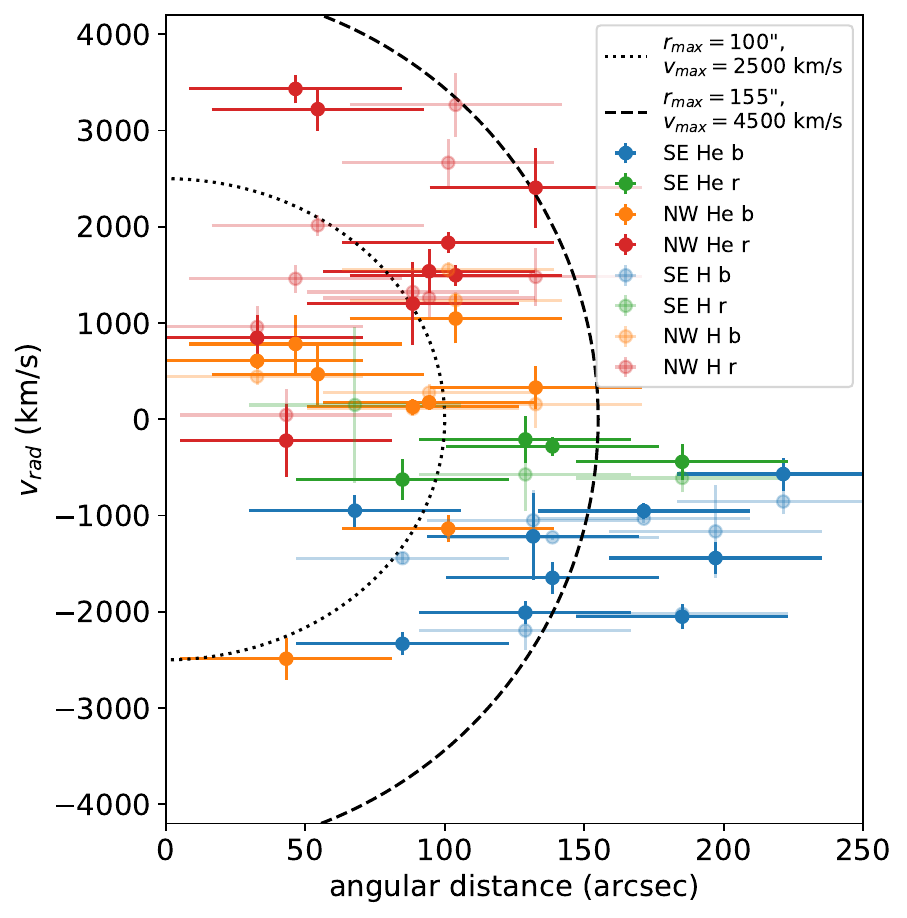}	
    }
  \caption{
  Radial velocities of all line components as a function of projected angular distance with
  respect to the explosion center determined by \citet{thorstensen01}.
   The horizontal error bars have a size of 38\arcsec, which is about half the angular resolution of the \xrism\
   telescopes. H-like line components have the same color coding as He-like lines, but are printed transparently.
   The date points with angular distance $>155$\arcsec, are mostly dominated by
    emission scattered by the telescope. The dashed line indicates the expected radial velocities for plasma heated by the forward shock. The dotted line indicates the expected radial velocities of reverse-shock heated plasma. 
    {Alt text:
    Scatter plot of measured radial velocities versus the angular distance from the center. Different colors show the division between He-like and H-like lines, as well as the difference between the red and blue components.
    }
    \label{fig:vrad_vs_dist2}
  }
\end{figure}

\begin{figure*}
\centerline{
   \includegraphics[trim=0 0 40 0 0,clip=true,width=0.5\textwidth]{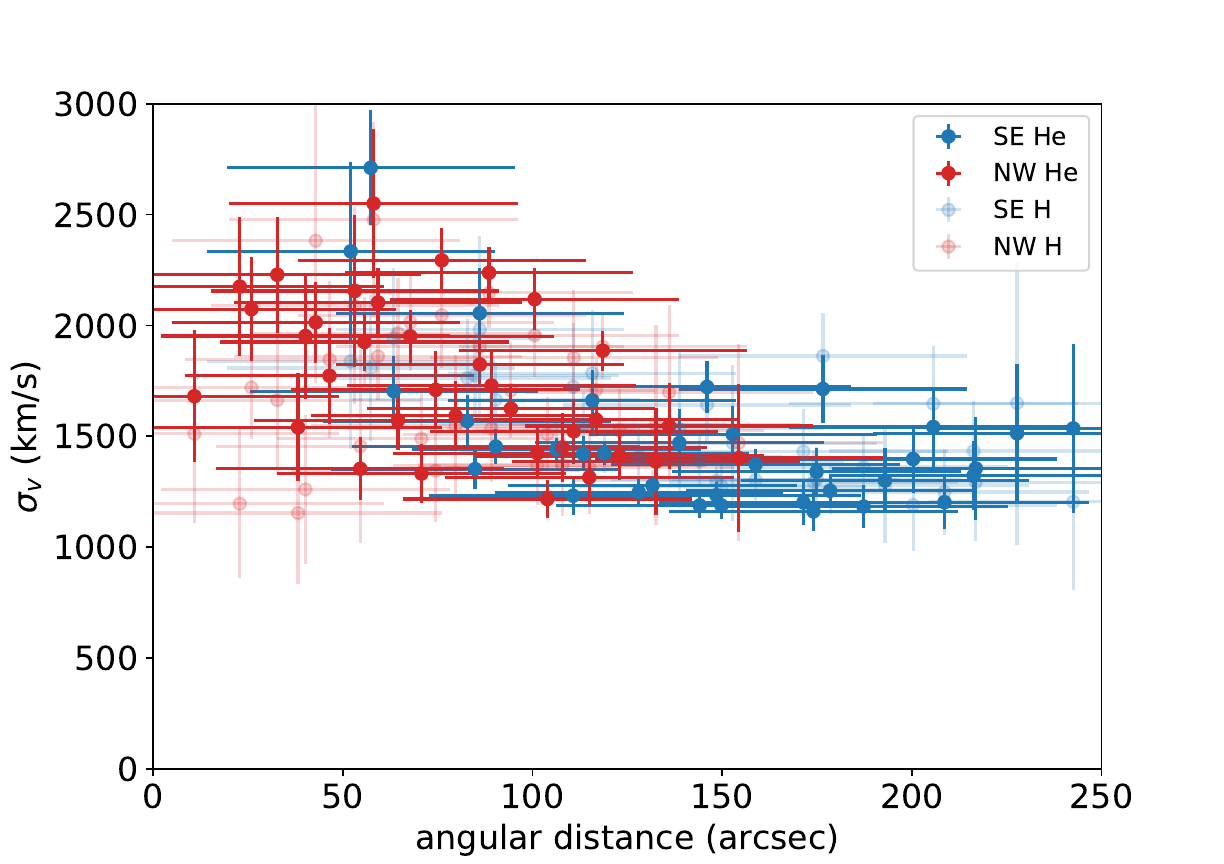}	
    \includegraphics[trim=0 0 40 0 0,clip=true,width=0.5\textwidth]{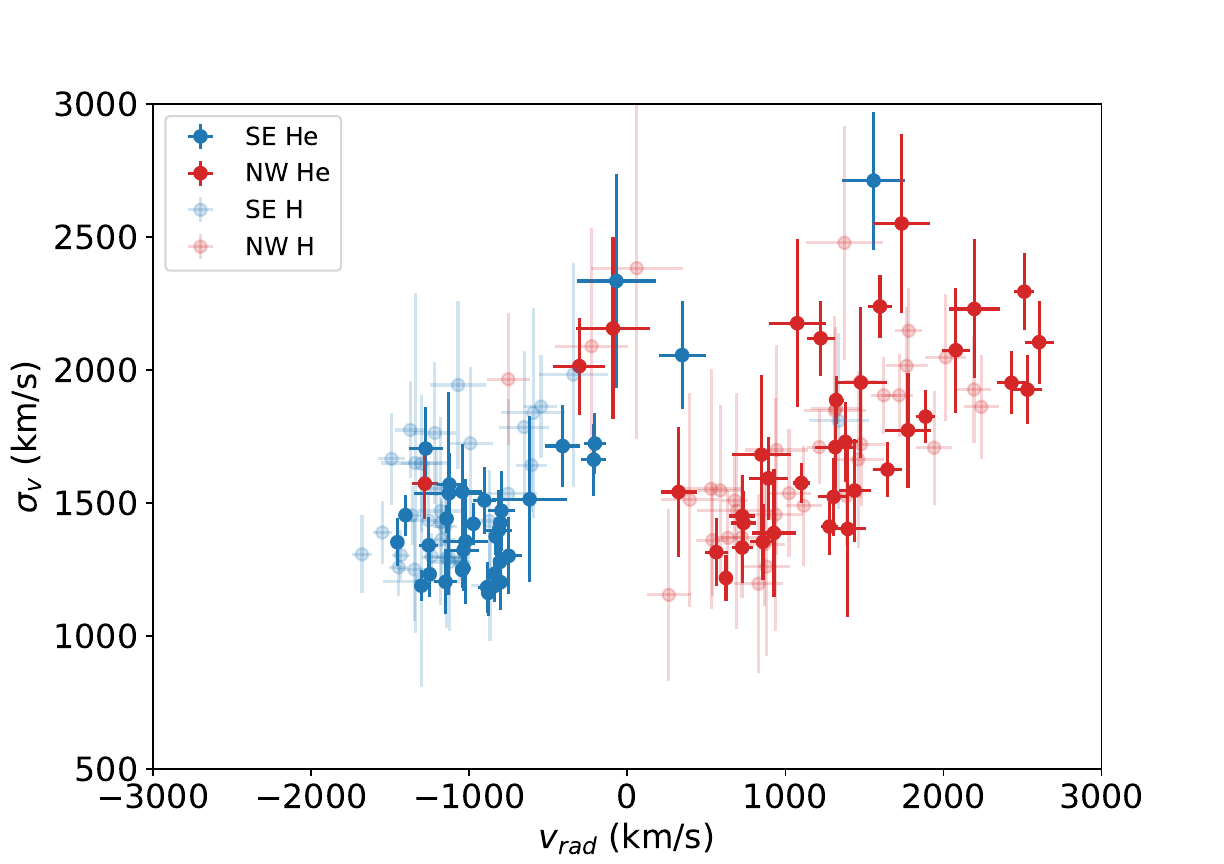}
    }
    \centerline{
       \includegraphics[trim=0 0 0 0 0,clip=true,width=0.5\textwidth]{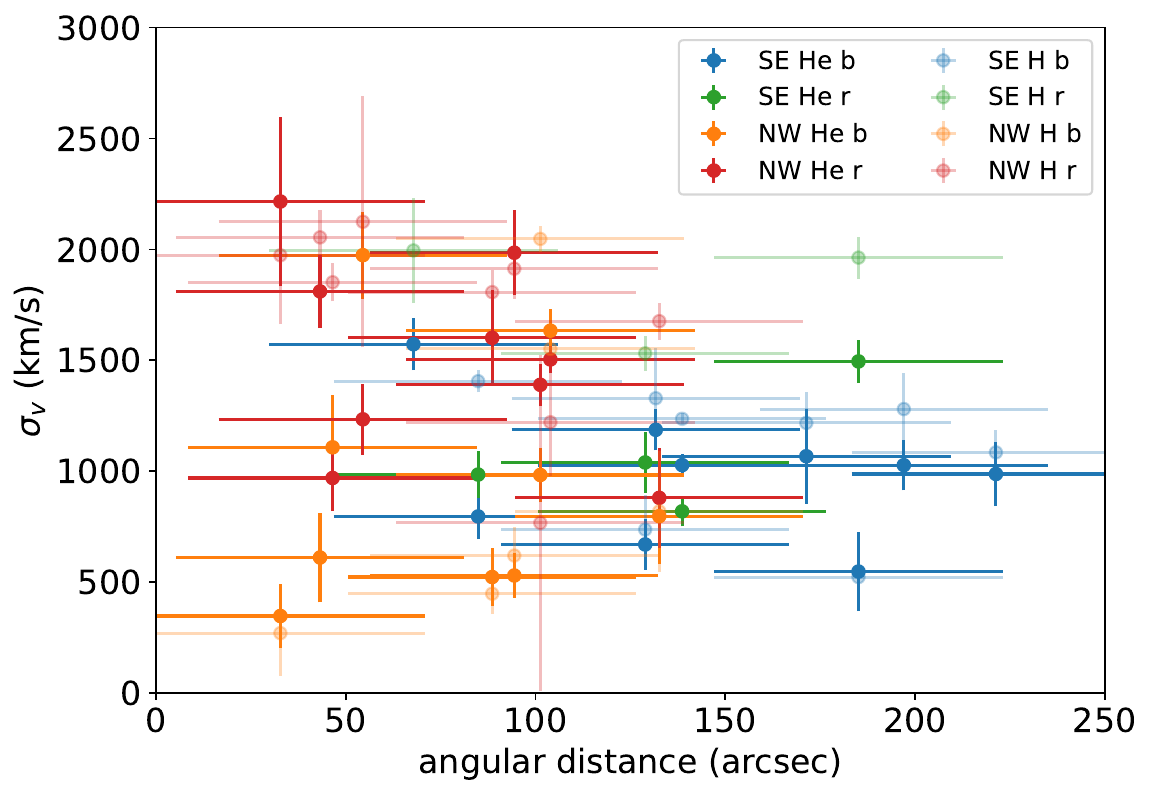}
        \includegraphics[trim=0 0 0 0 0,clip=true,width=0.5\textwidth]{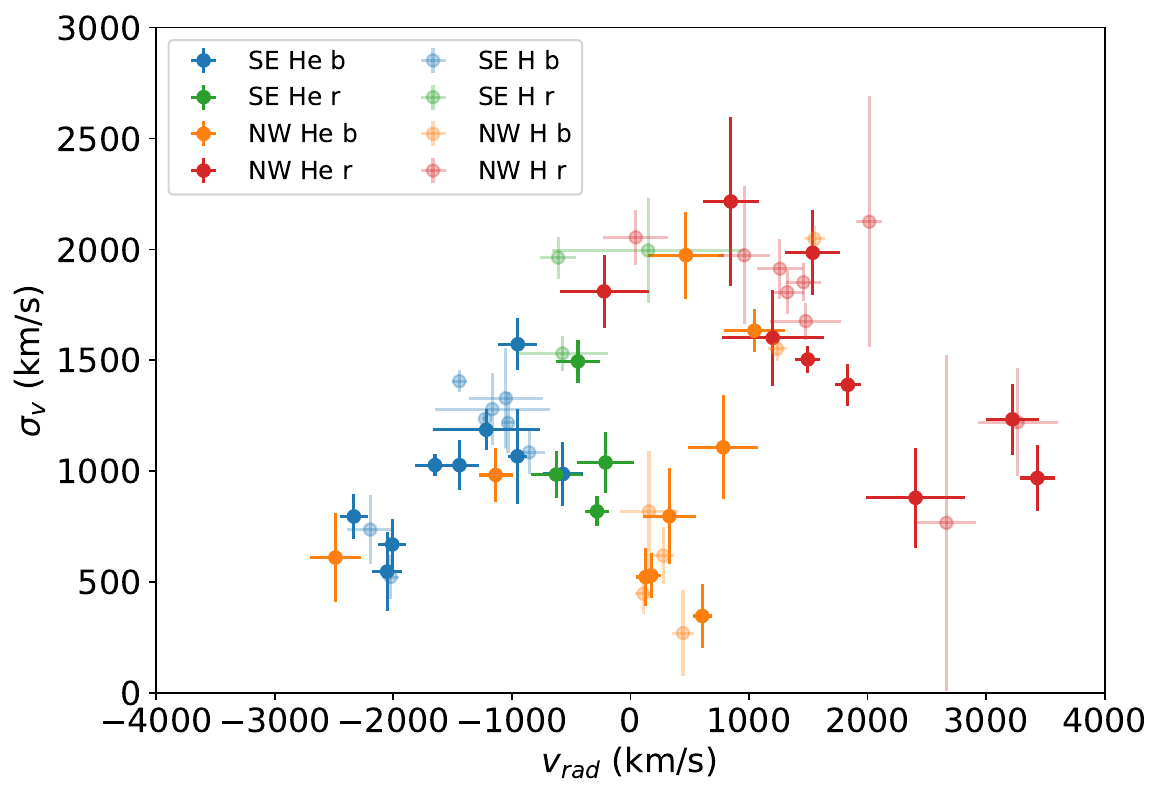}
    }
\caption{\label{fig:scatter_sigma}
Top row: scatter plots of $\sigma_v$ versus projected angular distance (left) and $\sigma_v$ versus the measured
Doppler velocity for single Gaussian component fits (see \sect~\ref{sec:automated}).
Bottom row: same as top row, but now for line spectral fitting using two Gaussian components.
H-like lines are shown in transparent colors.
{Alt text:
Four panels of scatter diagrams. Top left: velocity broadening versus angular distance for single components fits. Top right: velocity broadening versus radial velocity for single component fits. Bottom left: velocity broadening versus angular distance for two-component line fits, showing the separation between the blue and red components.
Bottom right: velocity broadening versus radial velocity for two-component line fits, showing the separation between the blue and red components.
}
}
\end{figure*}

The \xrism\ observations discussed here only cover the SE and NW parts of Cas A, about 60\% of the SNR, but they include almost all the brightest, ejecta-dominated parts of the
shell---see Fig.~\ref{fig:pixels}.
The SE/NW observations are also well adjusted to the blueshift/redshift dichotomy in X-ray line emission, first noted by \citet{markert83}. 

Surprisingly, the \xrism\ \resolve\ spectra show that in most regions the Si/S line emission does not consist  of a combination of  redshifted and blueshifted spectral components with one component dominating. What we report here is that if the line-shapes are best described by two components, rather than one broadened radial velocity component,  these components are either both redshifted (NW) or both blueshifted (SE), with the only exception being pixel {\it a} in the NW.
The \resolve\ spectra, therefore, indicate that 
the emission from the NW
 originates almost entirely from the backside of the SNR, whereas the SE part originates almost entirely from the frontside of the SNR.

These findings are  consistent with radio Faraday rotation measurements, which show that the rotation measure is larger in the NW than in the SE, indicating that
the radio emission from the NW traverses more plasma before reaching us \citep{kenny85}.  
Past X-ray Doppler measurements, however, did not address the relative fractions of front/backside emission for these regions.
With the \xrism-\resolve\ spectra we have been able to quantify the relative front/back asymmetry on spatial scales of $\sim 1$\arcmin.

In this paper, we focus on the radial velocities based on H-like and He-like silicon and sulfur lines in the 2-3 keV range, which we modeled using Gaussian line profiles
(\sect~\ref{sec:automated}).
This approach works well with the \xrism-\resolve\ spectra in the 2-3 keV band as the dominant lines are relatively isolated,  despite the intrinsic Doppler broadening of the lines.
However, this is no longer true for the Fe-K complex, where the range of charge states is broad, giving rise to closely packed lines, worsened by Cas A's intrinsic Doppler broadening.
In \citet{bamba25} the radial velocities are therefore fitted using a plane-parallel, non-equilibrium ionization (NEI) shock model.
Using Gaussian line components for the Si and S lines, as compared to NEI modeling, has the added advantage that line groups
like He-like and H-like lines can be treated completely independently.  We took advantage of \xrism-\resolve's ability to isolate individual lines by further modeling lines as being composed of two independent Gaussian components (\sect~\ref{sec:double}). See \citet{suzuki25} for a different approach to fitting the Doppler shifts for the He-like and H-like Si and S lines using NEI modeling.

In previous studies using CCD spectra  \citep{holt94,willingale02,hwang01b}, Doppler shifts were measured using 
line centroiding of unresolved line complexes. This makes CCD spectroscopy
vulnerable to energy shifts caused by ionization differences or other effects that are not related to Doppler shifts. 
Moreover, CCD spectroscopy does not allow one to judge whether a line-complex centroid shift is caused by bulk velocities, or
by having two Doppler shift components mixed with different contributions. 
In contrast to X-ray CCD spectroscopy,  the \xrism-\resolve\ instrument allows individual line shapes to be resolved. 
We explored this ability by fitting isolated lines using two Gaussian line components (\sect~\ref{sec:double}).
Perhaps surprisingly, the \resolve\ spectra reveal that the SE/NW  line energy dichotomy revealed by CCD spectroscopy turns out to be almost solely attributable to Doppler shifts.
Even more surprising is that even when attempting to fit the lines with two Gaussian components, we find that often both components are redshifted (NW) or  blueshifted (SE), but with one significant exception---NW pixel {\it a}, see Fig.~\ref{fig:vrad_maps_doublegauss}.

\subsection{Radial velocities}
\label{sec:vrad}
The Si and S X-ray line emission from Cas A is dominated by shocked ejecta. In order to interpret the three-dimensional distribution of the X-ray emitting ejecta---2 space dimensions plus radial velocity---it
is important to keep in mind  the 3D distribution of the optically emitting ejecta, consisting of the so-called
 fast moving knots (FMKs), which have been measured in more spatial detail than the X-ray emitting ejecta.
Note that the FMKs flare up when entering the reverse shock, which drives a radiative shock through their interiors, after which they are bright for $\sim$25~yr \citep{fesen11}.
The FMKs largely keep their original velocities after entering the reverse shock, unlike the diffuse X-ray emitting ejecta. As a result the optically bright FMKs show up near the current
location of the reverse shock, and their velocities are approximately their original free expansion velocities, i.e. $v_{\rm fmks}\approx R_{\rm rs}/t$, with $R_{\rm rs}$ the reverse
shock radius and $t\approx 350$~yr the age of Cas A.

The asymmetries in the velocities and locations of the ejecta revealed both in X-rays and optical are governed by i) asymmetries in the distribution of the ejecta as imparted by the supernova explosion, and ii)
asymmetries in the reverse shock radius, $R_{\rm rs}$.  The radius of the reverse shock is determined by several factors, including the velocity and density distribution of the ejecta, but
also the density structure of the CSM \citep[e.g.][]{chevalier82,truelove99,micelotta16,orlando22}.
The velocity distribution of  the FMKs in the northern part of Cas A shows that the
reverse shock is at a larger radius at the backside---with $v_{\rm fmks}\approx -5600$~\kms---than at the frontside of the SNR, with  $v_{\rm fmks}\approx 4000$~\kms\ \citep{reed95,milisavljevic13,alarie14}.

For a spherically symmetric ejecta distribution, the smaller the reverse shock radius, the more ejecta have been heated by the shock. 
If the ejecta distribution would indeed be spherically symmetric, one would, therefore, expect that 
the smaller reverse shock radius at the front side would have resulted in more hot ejecta, and hence, 
the X-ray emission would be dominated by the frontside of the SNR;
i.e. one would expect the X-ray ejecta to be mostly blueshifted.
The fact that the X-ray emission from NW is largely redshifted, despite a larger reverse shock radius at the backside, indicates that the asymmetries in the ejecta distribution in the NW part can only be explained by asymmetries in the
ejecta distribution (option i, mentioned above) as imparted by the supernova explosion, and not by the location of the reverse shock (option ii).

In the NW the 3D optical emission from FMKs \citep{lawrence95,milisavljevic13,alarie14} do show both redshifted and blueshifted components \citep[e.g. Fig. 6c, in][]{milisavljevic13}.
Indeed, for the two-Gaussian-component line models presented in \sect~\ref{sec:double} the NW is the only region where we find evidence for a truly blueshifted component, but only
in one region, NW pixel {\it a}, with a radial velocity of  $\sim -1150$~\kms  (Fig.~\ref{fig:vrad_maps_doublegauss}). 
This gives the impression that in the NW a blueshifted X-ray component is present, but is less prominent than the 
optical emission from the blueshifted FMKs. The blueshifted component in the NW reported here for pixel {\it a}, is consistent with blueshifted X-ray knots in that general region reported by \citet{lazendic06} based
on \chandra\ grating measurements, with velocities of $\sim -1500$~\kms\ and $-1700$~\kms. 

Note that in the optical and X-rays the northern part of the shell has changed quite dramatically over the last $\sim 70$~yr. For example, in Fig.~1 in \citet{patnaude14} one can see
that the lower part of the northern shell seems to have developed  since $\sim 1977$. This suggests that the FMKs in this location, which are indeed on the frontside \citep{milisavljevic13}, 
have only recently started entering the reverse shock ejecta. This affects the relative importance of the frontside emission in X-rays more than in the optical.
Since, the FMKs are typically bright for $\sim$25~yr, equal weight is given to the frontside and backside FMKs. However, the X-ray emission is determined by the total accumulated
ejecta that have been heated by the reverse shock, due to the very long cooling time of the plasma. For example, if relatively dense ejecta reached the backside starting $\sim 100$~yr ago,
and the frontside $\sim 40$~yr ago, the backside would dominate the emission. But since these timescales are longer than the emission timescales of FMKs, the front/back asymmetry
is less strong in the optical.

Although the 3D distribution of FMKs may be intrinsically different from the diffuse X-ray emitting ejecta, the Doppler shifts of both types of ejecta show a lack
of emission from the backside of the SNR in the SE:
the \xrism-\resolve\ spectra confirm earlier results \citep{markert83,holt94,willingale02,hwang02} that the SE emission is overall blueshifted. Moreover, modeling the line emission with two Gaussian components did not reveal any significant evidence for a truly redshifted component in the SE.
This matches with 
the 3D reconstructions of the FMKs as presented in \citet{lawrence95,milisavljevic13,alarie14}, which show no evidence for
redshifted FMKs in the SE of the SNR \citep[e.g. Fig. 6c, in][]{milisavljevic13}.

Comparing the optical Doppler measurements to the X-ray Doppler data reported here reveals that the optical knots probe a much broader range in radial velocities.
This can be attributed mainly to the different dynamics of the optically emitting FMKs. Since these are dense knots,
they have hardly been decelerated by the reverse shock.
Hence, the optical knot velocities are close to the free expansion velocity of the ejecta.
The X-ray emitting ejecta, on the other hand, are ejecta shocked by the reverse shock, which heats the plasma, but also slows it down to a velocity of $\Delta v = \frac{1}{4} V_{\rm rs}$ with respect to the reverse shock velocity, $V_{\rm rs}$, in the unshocked-ejecta frame.  The prefactor  comes from $\frac{1}{\chi}=\frac{1}{r}$,
with $\chi=4$ the compression ratio for a strong shock heating a monatomic gas.
The reverse shock velocity in the unshocked-ejecta frame is the relative velocity between the reverse shock displacement velocity $dR_{\rm rs}/dt$ (observer's frame) and the free expansion velocity of the unshocked ejecta, i.e.
\begin{equation}
V_{\rm rs}= \frac{R_{\rm rs}}{t} - \frac{dR_{\rm rs}}{dt}. \label{eq:v_rs}
\end{equation}
The shocked-ejecta plasma velocity in the frame of the observer is then
\begin{equation}
v_{\rm ej}= \frac{dR_{\rm rs}}{dt} + \Delta v_{\rm ej}  
= \frac{3}{4} \frac{dR_{\rm rs}}{dt} + \frac{1}{4}\frac{R_{\rm rs}}{t} = \frac{1}{4}\left( 3m + 1\right)\frac{R_{\rm rs}}{t}, \label{eq:v_ej}
\end{equation}
where we defined the expansion parameter $m$ by $dR_{\rm rs}/dt = m R_{\rm rs}/t$, 
and used $\Delta v_{\rm ej}=\frac{1}{4}V_{\rm rs}$, with $V_{\rm s}$ given by (\ref{eq:v_rs}).

According to SNR evolutionary models for Cas A $dR_{\rm rs}/dt \sim 1000$--$2000$~\kms\ \citep{orlando22,micelotta16}.
However, X-ray expansion measurements reveal that there is likely a large variation in $dR_{\rm rs}/dt$.
In particular
in the western part the reverse shock is moving toward the center, down to velocities of  $dR_{\rm rs}/{dt}\approx -2000$~\kms\ \citep{helder08,sato18,vink22a}, probably in response to high-pressure build up
when the forward shocked started interacting with a dense CSM structure
\citep{orlando22}.
The reverse shock radius in Cas A is $1.9\pm 0.2$~pc \citep{arias18}, so that $R_{\rm rs}/t\approx 5300\pm 550$~\kms---in agreement with the range of velocities for the optical knots. 
However, with $-0.38 \lesssim m \lesssim 0.7$,
we can expect ejecta plasma velocities of $ -185 \leq v_{\rm ej} \lesssim 4500$~\kms, but with a typical velocity of 2450~\kms, assuming $dR_{\rm rs}/dt = 1500$~\kms and $R_{\rm rs}/t=5300$~\kms.

Indeed, the X-ray Doppler velocities seem to encompass that range. For the spectral fits with single Gaussians we see that the radial velocities in the SE encompass a smaller range of $|v_{\rm rad}|\lesssim 2000$~\kms\
than for NW,  $|v_{\rm rad}|\lesssim 3000$~\kms\ (Fig.~\ref{fig:hist}). For the spectral fits with two Gaussian components the range for the SE is somewhat larger,
$|v_{\rm rad}|\lesssim 2500$~\kms, but still less than for the NW part: $|v_{\rm rad}|\lesssim 3500$~\kms---see Fig.~\ref{fig:vrad_hist_2gauss_separated}. The different maximum Doppler velocities in the SE and NW probably reflect the front-back asymmetry of the optical knots,
with the backside (redshifted components) having a larger reverse shock radius.

Apart from this asymmetry in maximum velocities, most components with the highest $|v_{\rm rad}|$
have  radial velocities consistent with the broad range of expected ejecta plasma velocities
indicated by Eq.~\ref{eq:v_ej}.
We illustrate this in Fig.~\ref{fig:vrad_vs_dist2}, which shows the radial velocities obtained from the double Gaussian fits as a function of projected radius.
The dotted and dashed line show expectations for two choices of maximum radius, with $r_{\rm max}=100$\arcsec\ (1.67~pc) corresponding to the minimum value of reverse shock radius, and $r_{\rm max}=155$\arcsec\ (2.6~pc) corresponding to the forward-shock radius. 
The components with the highest $|v_{\rm rad}|$ are shown in blue (SE) and red (NW), and
fall in general within the expected range of radial velocities indicated by the dashed and dotted lines.
Exceptions are the red points within 50\arcsec\ of the center. 
For the SE pointing most of the less extreme radial velocity components (faint green points) are
within the expected boundaries. However, there are many low radial velocity components
associated with the lower radial velocities of the NW pointing (faint orange points)
that are not within the expected boundaries. So there are a number of components that are not easily associated with the plasma velocities expected for reverse-shock heated ejecta, or forward shock
heated plasma. These components are associated with line emission from the central region, and
also stand out in their $\sigma_v$ properties, as discussed below.

We have to be cautious not to overinterpret these points, but there are a few possibilities as to how the radial velocities toward the projected center could indeed be that low.
First, as already mentioned, a reverse shock that moves toward the center ($m=-0.38$) can even impart (small) negative velocities to the shocked ejecta. 
Second, it is possible that toward the center we indeed do encounter both the front and backside of the SNR shell with roughly equal emission rates, and the
two components cancel each other out. This does not seem very likely as this would lead to a very broad line, which would be appropriately captured by
a very negative and positive radial velocity in our fits with two Gaussian components. This is not at all what
is measured, see \sect~\ref{sec:sigma_v}

Finally, we also have to consider the possibility that some line emission originates from shocked CSM. For the forward shock the plasma velocity is simply given by $v_{\rm csm}=\frac{3}{4}V_{\rm fs}\approx  4500$~\kms, for $V_{\rm fs}\approx 6000$~\kms\ \citep{vink22a}. The range of line-of-sight velocities for the forward shock are depicted by the dashed line in Fig.~\ref{fig:vrad_vs_dist2}.
 Recently \citet{vink24a} reported based on \chandra\ data  that most of the interior emission 
originates from shocked CSM. In particular, the JWST-discovered  structure dubbed the ``green monster" \citep{milisavljevic24,delooze24} is a CSM structure, with a relatively low
blueshift, suggesting a dense CSM structure that may have slowed down the forward shock velocity. However, the  ``green monster" appears to have radial velocities of $\sim -2300$~\kms, which is slow, but still
much faster than the low-velocity components at small angular distances seen in Fig.~\ref{fig:vrad_vs_dist2}.

\begin{figure*}
  \centerline{
    \includegraphics[trim=50 50 100 80,clip=true,width=0.5\textwidth]{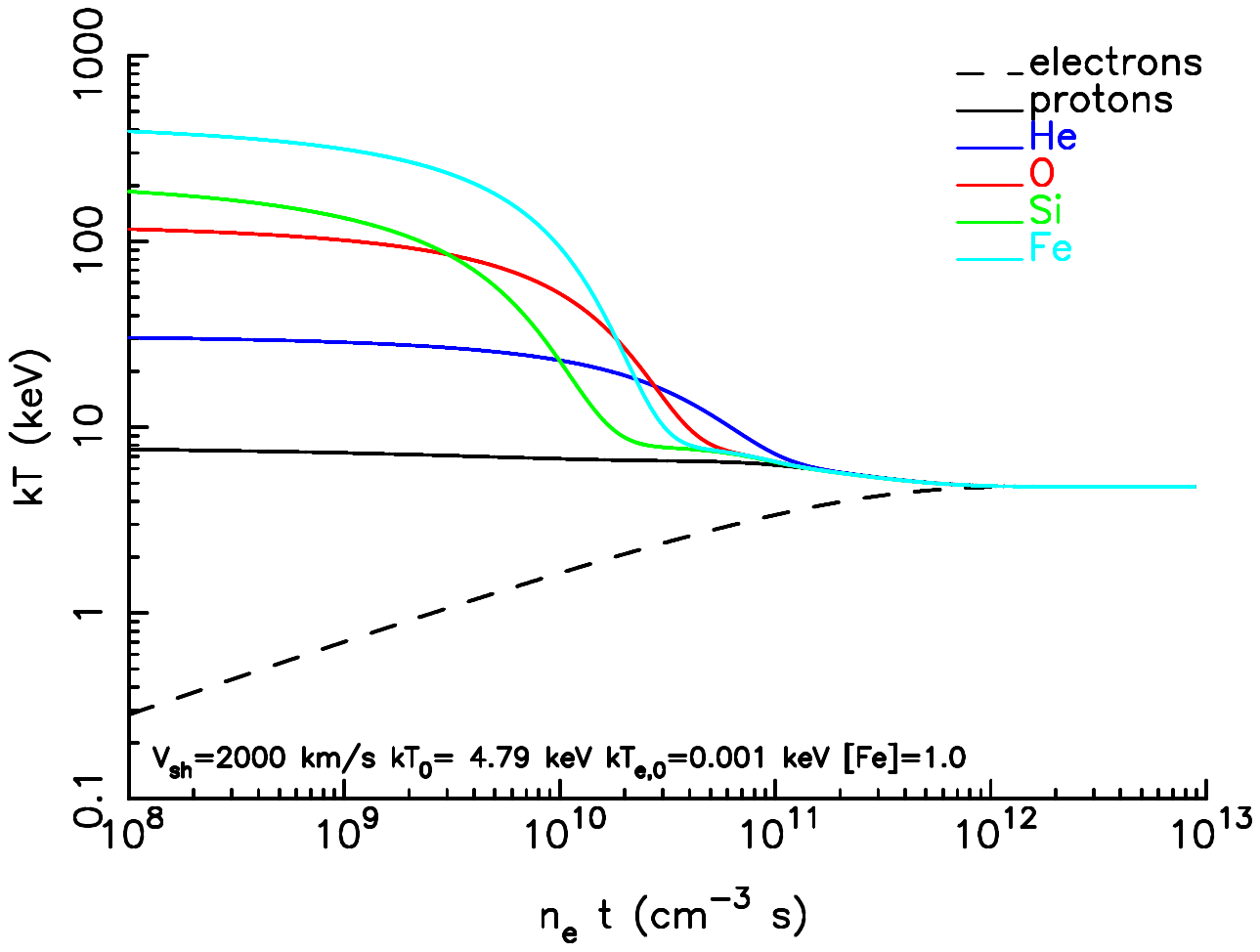}
    \includegraphics[trim=50 50 100 80,clip=true,width=0.5\textwidth]{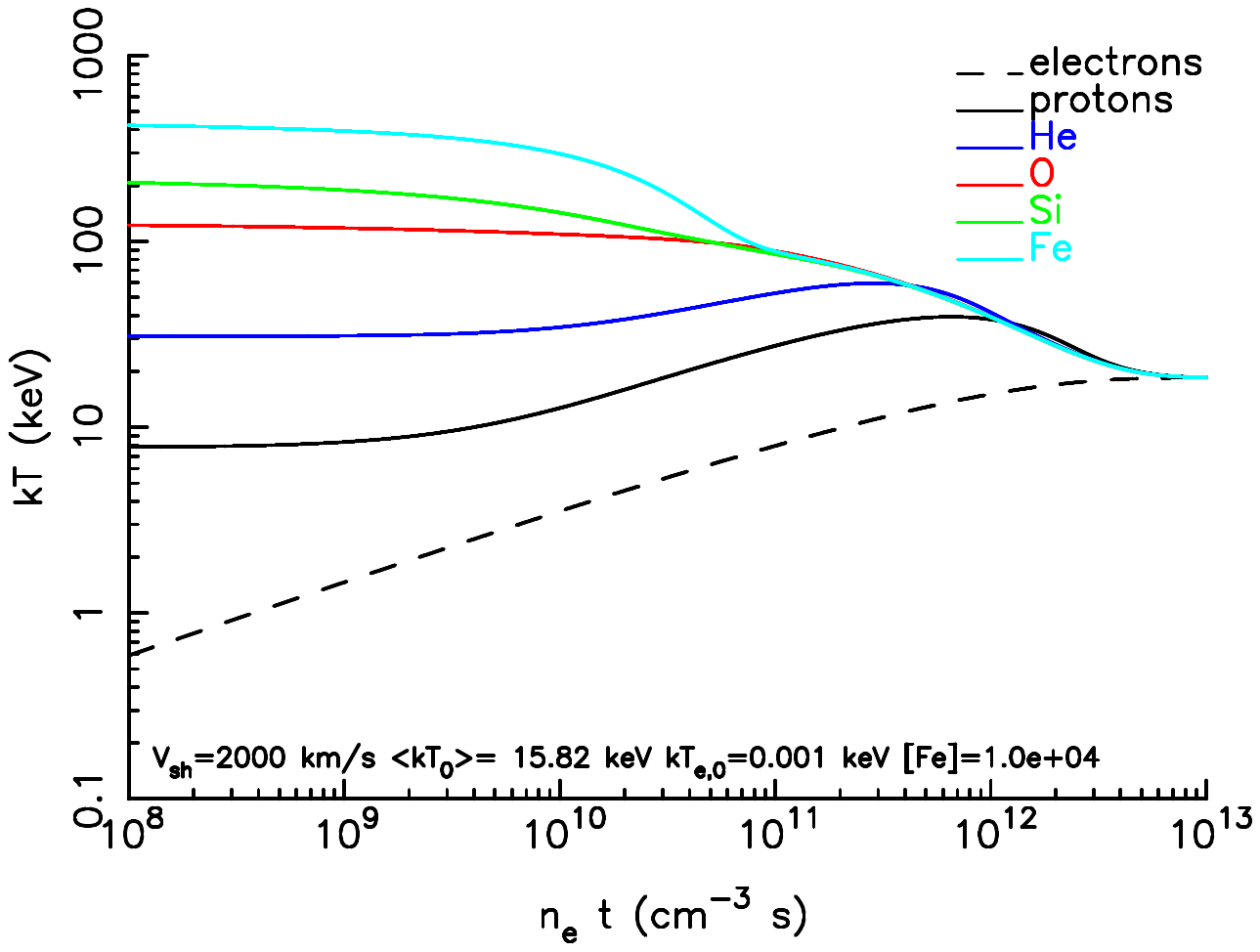}
  }
  \caption{
    Two plots that show the level of temperature equilibration for electrons,
    protons, helium, oxygen, silicon and iron ions as a function of
    $n_{\rm e}t$ for a shock velocity of 2000~\kms. Left: the case of a solar metalicity plasma.
    Right: the case for a proton-, helium-deficient plasma, as to be expected for the ejecta plasma in Cas A. For illustrative purposes the O, S, and Fe abundances have been set to $10^4$ times solar abundances.
    {Alt text: two panels showing the evolution of the ion and electron temperatures as a function of electron-density times time. The assumed shock velocity is 2000 km/s. The left panel shows the case for solar abundance plasma. The right panel shows the case for pure ejecta plasmas.}
    \label{fig:equilibration}
  }
\end{figure*}

\begin{figure}
\centerline{
\includegraphics[width=\columnwidth]{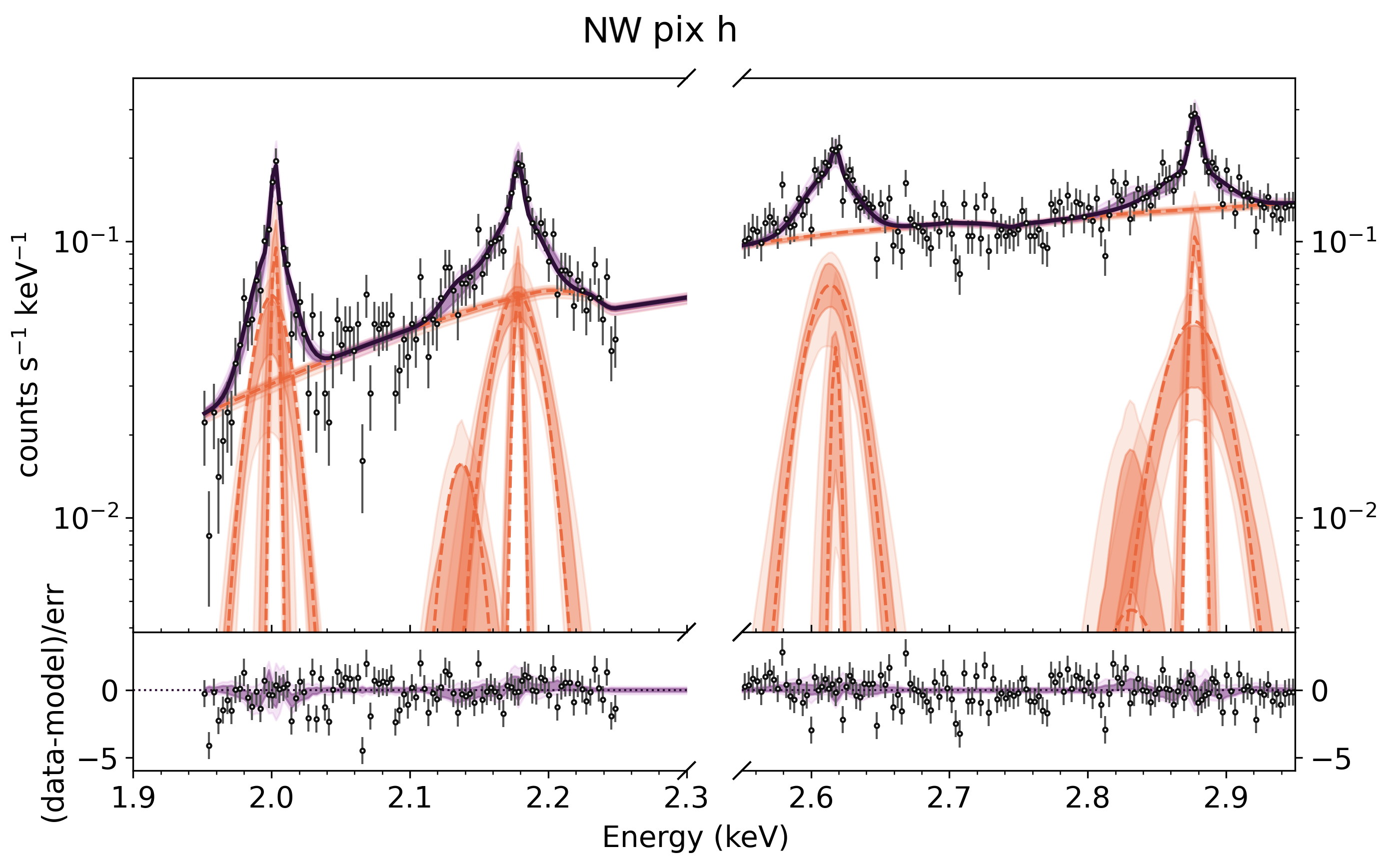}
}
\caption{
Fit results of NW pixel {\it h} with two Gaussian components. 
This pixel covers a part of the center of Cas A.
It shows that there is a narrow component ($\sigma_v\approx 300$~\kms) with a small redshift corresponding
to $v_{\rm rad}\approx 500$~\kms, which results in unusual, ``spiky", line profiles for all
four lines.
{Alt text: example of a spectrum with very spiky profiles (NW pixel h).}
\label{fig:pixh}
}
\end{figure}

\begin{figure}
\centerline{
\includegraphics[width=\columnwidth]{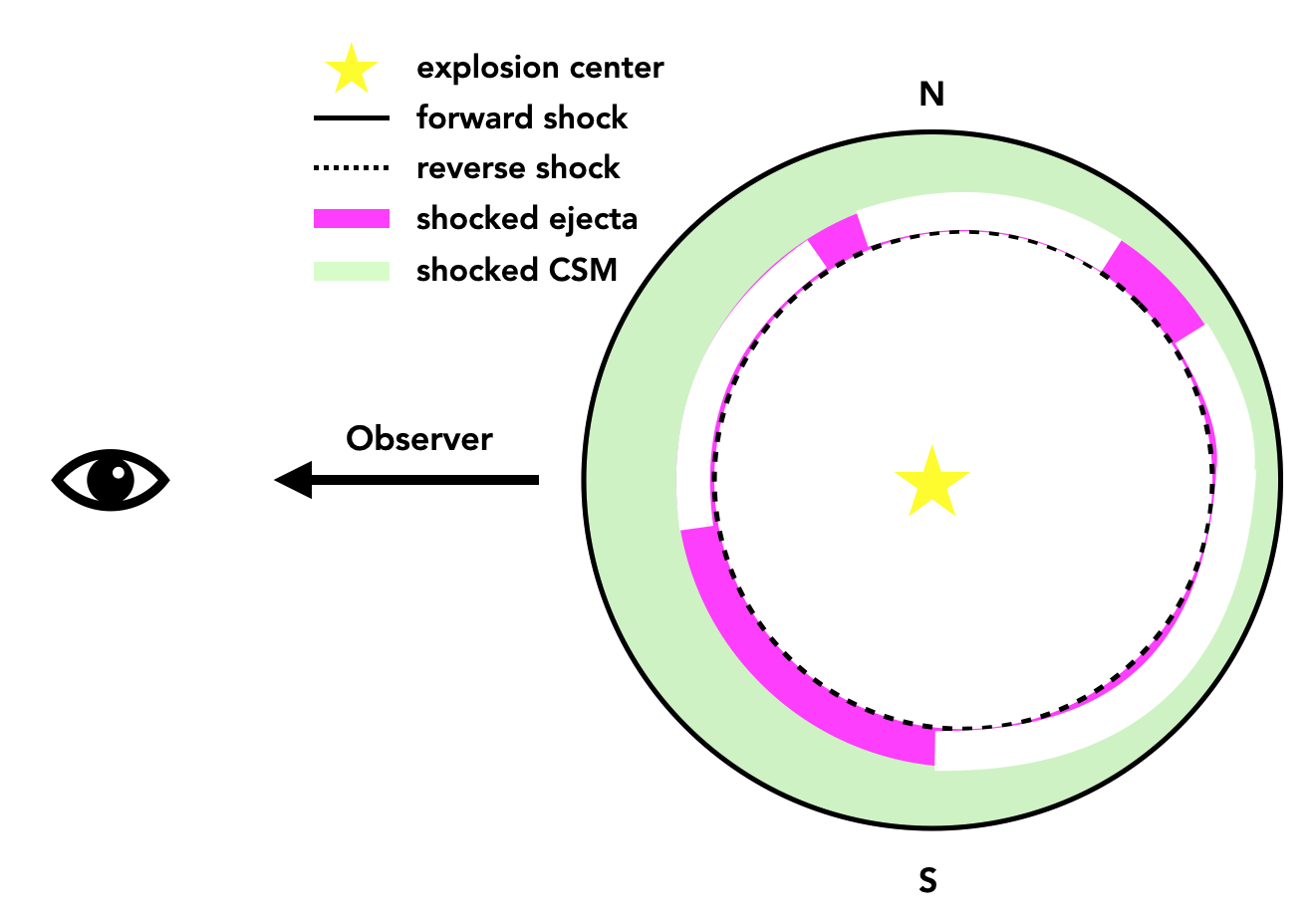}
}
    \caption{
    Cartoon of a ``sideways" look at Cas A. The reverse shock (dotted line) is displaced with respect to the explosion center (yellow star), accounting for the higher ejecta velocity at the backside of the shell. The line emission comes mostly from the shocked ejecta (magenta), but also from the shocked CSM (light green).  Shocked CSM dominates the X-ray emission toward the center of the SNR.
    {Alt text: cartoon showing a side view of Cas A, indicating the locations of shocked ejecta and shocked CSM.}
    \label{fig:cartoon}}
\end{figure}

\subsection{Line broadening}
\label{sec:sigma_v}

The Cas A spectra show severe intrinsic line broadening of typically
$\sigma_v\sim 1500$~\kms, corresponding to $\Delta E=12.5$~eV at 2.5~keV. This is to be compared to
the \resolve\ resolution of $\Delta E\approx 5$~eV. 

The origin of this line broadening is a combination of Doppler broadening due to line of sight velocity distributions (macro movements), and thermal velocity broadening of the ions (micro movements). 
We attempted to quantify the line of sight motions by fitting the lines with two Gaussian components, assuming the line of sight motions can be decomposed into two large velocity components. However, in most cases these two components
cannot be attributed to the frontside and backside of the SNR.
Irrespective of exact geometry,
Doppler broadening due to  line-of-sight motions are expected to be the largest in the central region, where $v_{\rm rad}=v \cos\theta$ is maximized,
with $\theta$ the inclination angle. As Fig.~\ref{fig:maps_Gaussians} shows, the line broadening indeed peaks in the central regions. But as Fig.~\ref{fig:scatter_sigma} (top left) shows $\sigma_v$ does not peak at the very center $<50$\arcsec, but
near $70$\arcsec.  It does show that for large projected radii $\sigma_v$ lowers to $\sigma_v\approx$1200--1600~\kms.
Since for large angular radii line of sight motions are minimal, one could tentatively associate $\sigma_v\approx$1200~\kms\
to thermal broadening.
Fig.~\ref{fig:scatter_sigma}  also shows that for some central regions the H-like emission is surprisingly low,
$\sigma_v\approx 1200$~\kms. These low $\sigma_v$ measurements correspond to the equally surprising low $v_{\rm rad}$
near the center discussed in \sect~\ref{sec:vrad}.

The radial velocity measurements indicate that the ejecta shell of Cas A consist of a patchy shell,
with the mostly backside of the shell emitting X-rays in the NW region, and for X-ray emission from  the
SE region only the frontside appears to be responsible. 
Since $v_{\rm rad}$ shows peak for the central regions in such
a system, and because the central region should contain also the largest gradients in $v_{\rm rad}$, we expect
a positive correlation between $\sigma_v$ and $|v_{\rm rad}|$. Fig.~\ref{fig:scatter_sigma} (top right) shows that
this is roughly the case for the NW region. However, for the SE region there is roughly an anti correlation
between  $\sigma_v$ and $|v_{\rm rad}|$. One could understand this if the shell is more thin for the high radial
velocity components. But it is not a priori clear why the NW and SE show a different behavior. 

The different behavior between the SE and NW disappears if one plots $\sigma_v$ versus $v_{\rm rad}$ for
all the significant individual velocity components measured assuming two Gaussian components (Fig.~\ref{fig:scatter_sigma},
bottom right). We see that then both the SE and NW have an anti-correlation between $\sigma_v$ and $v_{\rm rad}$,
but with the exception of several data points with $|v_{\rm rad}|\lesssim 1000$~\kms\ and $\sigma_v\lesssim 1000$~\kms.
These points again correspond to the low radial velocity components discussed before.
Apart from these points, the fact that the SE and NW points show up more symmetrically adds to the plausibility of
the two component Gaussian fits.

Fig.~\ref{fig:scatter_sigma} (bottom left) shows that the maximum $\sigma_v$ as function of projected radius
exhibits the expected trend that $\sigma_v$ declines with radius, but still at any radius
we find also components with low $\sigma_v$ ($\lesssim 750$~\kms), even as low as $\sigma_v\approx 300$~\kms. But the lowest radial velocity components appear more centrally located.

Before discussing the low $\sigma_v$ components toward  the center, we first put the line broadening in the context
of the expected thermal line broadening. 
Thermal Doppler broadening at the level of the $\sigma_v$   reported  here is only expected if the  shock heats all particle species independently. It is now well established that for collisionless shocks of young SNRs the electron and ion temperatures are indeed not equilibrated at the shock \citep[e.g.][]{ghavamian13,vink15}.
See also the recent \xrism\ results for SNR N132D \citep{xrism_n132d}.
In the most extreme case each species $i$ will be heated by a high-Mach-number shock to 
\begin{equation}
    kT_i=\frac{3}{16}m_i V_{\rm s}^2 \label{eq:kTi},
\end{equation}
with $m_i$ the mass of the particles. For a fully temperature equilibrated plasma all species have 
$kT_i=\frac{3}{16} \mu m_{\rm p} V_{\rm s}^2$, with $\mu m_{\rm p}$ the average particle mass---$\mu\approx 0.6$ for solar abundance plasmas.

Post-shock equilibration through Coulomb collisions proceed at time scales given by $n_{\rm e}t\approx 10^{10}$--$10^{13}$~\netunit,
as illustrated in Fig.~\ref{fig:equilibration}.
As the equilibration depends on the mass ratios of the species and the charge of the particles, full equilibration of electrons and ions tends to be longer for metal dominated plasmas (Fig.~\ref{fig:equilibration}, right). 
Moreover, for metal-dominated plasmas the final equilibrated temperature will be larger due to the larger average mass $\mu m_{\rm p}$.
The thermal Doppler line broadening is given by $m_i \sigma_{v,i}^2=kT_i$. Combining this with Eq.~\ref{eq:kTi} gives for the fully non-equilibrated case
\begin{equation}
\sigma_{v,i} = \frac{\sqrt{3}}{4}V_{\rm s}\approx 0.43 V_{\rm s}.\label{eq:sigma_vs}
\end{equation}
We can, therefore, directly infer from thermal broadening the shock velocity, which for the reverse shock is given by Eq.~\ref{eq:v_rs}.

The question now is what fraction of the measured 
$\sigma_v$ is associated with thermal line broadening. As discussed above at the outer
radius of the SNR one expects the thermal broadening to dominate over bulk radial velocity broadening.
Since, some of the line-sight-broadening was taken out by allowing for two velocity components, we resort
to the two Gaussian component fits. 
Fig.~\ref{fig:scatter_sigma} (bottom left) shows that at large projected radii we find $\sigma_v\approx 1000$~\kms,
although with one component measuring as high as $\sigma_v \approx 2000$~\kms and two components measuring as low as $\sigma_v \approx 500$~\kms.

Taking $\sigma_v\approx 1000$~\kms, which also provides a lower limit to $\sigma_v$ for single component fits, translates into 
$V_{\rm rs} \approx 2300$~\kms. For a free expansion velocity of $R_{\rm rs}/t \approx 5300$~\kms, this implies
a reverse shock displacement velocity of $dR_{\rm rs}/dt \approx 3000$~\kms, see Eq.~\ref{eq:v_rs}.
This is relatively large, since we expected $dR_{\rm rs}/dt \approx 1500$~\kms, which would give $\sigma_v\approx 1600$~\kms.
Reverse shock displacements with zero velocity would result in $\sigma_v\approx 2300$~\kms, which is still consistent
with at least one measured $\sigma_v$ at large radius.

So it is possible that the range of $\sigma_v$ at large radii reflect a range in reverse shock velocities in the
regions covered by the \xrism-Resolve field of view. Moreover, the relatively low typical value of $\sigma_v\approx 1000$~\kms,
corresponds to $kT_i\approx 280$~keV for Si. But the Si/S ion temperature may already have partially equilibrated
with the other ions, mostly notably oxygen, given that the ionization age in Cas A is typically $n_{\rm e}t\approx 10^{11}$~\netunit\ \citep[e.g.][]{hwang12}. See Fig.~\ref{fig:equilibration}.
So the derived $V_{\rm rs}$ for a given thermal $\sigma_v$ is likely an upper limit, and the $\sigma_v$
at large radii are consistent with the range of reverse shock velocities expected, if one takes into account
partial post-shock equilibration.

This leaves still the small values of $\sigma_v\lesssim 500$~\kms\ to discuss. If these are due to thermal
broadening, they correspond to $V_{\rm s} \lesssim 1200$~\kms, which is smaller than one would expect from
models of the reverse shock evolution of Cas A. They are also lower than can be accounted for by partial equilibration
in a metal-dominated plasma. More likely is it that they correspond to partially equilibrated plasma
of the CSM, for which the equilibration of all ion species with protons and helium occurs for $n_{\rm e}t\lesssim 5\times 10^{10}$~\netunit\ (Fig.~\ref{fig:equilibration}, left) for $V_{\rm s}\approx 2000$~\kms.
So most likely these narrow components are associated with Si/S line emission originating from shocked CSM.
There are, however, a few problems with that. First, the forward shock velocity in Cas A is $\approx 6000$~\kms,
which results in a hotter plasma temperature, and longer  ion equilibration times of $n_{\rm e}t\approx 5\times 10^{11}$~\netunit. 
But in particular the narrow components near the center of the SNR are problematic, as they have a small
radial velocity of $v_{\rm rad}\lesssim 500$~\kms. Their central locations do suggest that these components indeed originate
from the CSM \citep{vink24a}. However, in the center we expect either a very broad, or double peaked, line profile, which we
do not observe, or we expect a very blue- or redshifted component.  Instead, we observe that for some
NW pixels, the line shape is described by a broad component along with a narrow component with
a relatively small radial velocity. They show up as orange points with $\sigma_v\lesssim 750$~\kms\
and $|v_{\rm rad}|\lesssim 750$~\kms\ in Fig.~\ref{fig:scatter_sigma} (lower right).
The corresponding pixels are {\it f, h} and {\it i}. As an example, we depict
in Fig.~\ref{fig:pixh}, pixel {\it h}. Inspecting the line profile visually shows that
the result is very solid, as all four lines show the same spiky profile.

In Fig.~\ref{fig:cartoon}, we show a cartoon illustrating the ejecta and CSM morphology of Cas A inspired by
the X-ray and optical Doppler mapping. It shows a situation in which the central regions are dominated by
shocked CSM, but it should result in a combination of highly redshifted and highly blueshifted plasma. 
It is not clear what could be the location of low radial velocity plasma. A possible explanation could be that it originates from very dense material, which is overrun by the shock, which then rapidly slows down. But this should give
rise to either bright clumps or regions, with shocks rapidly evolving into radiative shocks. There is no X-ray or optical
evidence for such a component.

The Doppler velocities and broadening confirm the notion that the ejecta shell is patchy. It is tempting to
associate the SE/NW velocity dichotomy as being evidence for a bipolar explosion, but the detailed \resolve\ line fits, as well as the 3D distribution of FMKs, rather indicate an irregular shell of ejecta. The current \xrism\ pointing do not cover the locations that are likely connected with a bipolar explosion component: the ``jets". We do note, however, that some of the highest line broadening off-center are found at the base of the NE jet (SE pixel 11, or {\it a}) for the H-like ions. We speculate that this may be related to a very hot ion temperature, caused by having been shocked with a very high shock velocity, associated with the passage of the NE jet.

In conclusion, although the \xrism-\resolve\ radial velocity measurements confirmed the overall SE/NW blueshift/redshift
dichotomy, the detailed line fits show an unexpected feature that for now defies an easy explanation.

 \section{Conclusions}
With the launch of \xrism\ the era of high-spectral resolution imaging X-ray spectroscopy has finally arrived. This allows us to make accurate radial velocity maps of extended X-ray sources. We reported here a radial velocity study of the young core-collapse SNR Cas A, using the Si and S lines in the 2-3 keV range. With the high spectral resolution of the
\resolve\ instrument of  $\Delta E\approx 5$~eV individual He-like and H-like lines can be fitted, and their line
shapes can be modeled.

Although the radial velocity maps that we reported here agree in general with the overall redshifted and blueshifted structure determined through CCD X-ray spectroscopy, the high-resolution spectra from \xrism\ have revealed many more details in these complicated spectra.  The spectral analysis results in many interesting and puzzling findings:
\begin{itemize}
\item With CCD spectra one cannot make a distinction between a mix of blueshifted and redshifted spectra. Detailed line model of the \resolve\ spectra show that only in a few cases the line of sight encounters a combination of blueshifted and redshifted plasmas, and only in the NW region.
\item Decomposing the line shapes in two components in most cases result in two components that are both redshifted, or both blueshifted.
\item The radial velocity distribution reinforces the idea from optical imaging spectroscopy that the ejecta shell is very patchy, with no backside component present in the SE, and only a small frontside component is present in the NW.
\item The intrinsic line broadening suggest an average thermal line broadening of $\sigma_v\approx 1000$~\kms, but with some variation consistent with different reverse shock velocities, and  partial equilibration.
\item There are puzzling velocity components projected toward the center of Cas A with a very low
line broadening $\sigma_v\lesssim 500$~\kms\ and small radial velocity of $|v_{\rm rad}|\lesssim 500$~\kms.
These components are likely associated with shocked CSM, but the low radial velocities in combination
with a location projected toward the center of Cas A cannot be easily interpreted.
\end{itemize}

The analysis presented here shows the power of high-resolution imaging spectroscopy, and for Cas A reveals the importance
modeling both the ejecta and the CSM components. For a better understanding of Cas A it is important to cover
also the rest of Cas A with \resolve. In particular not observed yet are the SW region, which covers
a region where the forward shock encountering an arc of dense CSM material, which may provide some insights also into the puzzling CSM component projected toward the center. In the western region the reverse shock is moving inward, resulting in 
a strong reverse shock velocity. And finally in the NE region the jet is prominent feature. 
Complete coverage will, therefore, provide more insights in the origin of the different velocity components revealed
in the SE and NW pointings reported here.

\begin{ack}
 We thank Masahiro Tsujimoto for his constructive comments as an internal XRISM reviewer. This helped to improve
 the quality and clarity of the manuscript.
\end{ack}

\section*{Funding}
 The work on this project by JV and MA was (partially) funded by NWO under grant number 184.034.002.
 This work was  supported by the JSPS Core-to-Core Program (grant number: JPJSCCA20220002), and the Japan Society for the Promotion of Science Grants-in-Aid for Scientific Research (KAKENHI) Grant Number JP23H01211 (AB), JP23K25907 (AB), and JP20K04009 (YT).

\section*{Data availability} 
The spectral and analysis data underlying this article will be made available through \url{zenodo.org} under
\url{https://doi.org/10.5281/zenodo.14734013}.

\appendix %%%%%%%%%%%%%%%%%%%%%%%%%%%%%%%%%%%%%%%%%%%%%%%%%%%%%%%%

\section*{Appendix. Supplemental figure}
\begin{figure*}
\includegraphics[width=\textwidth]{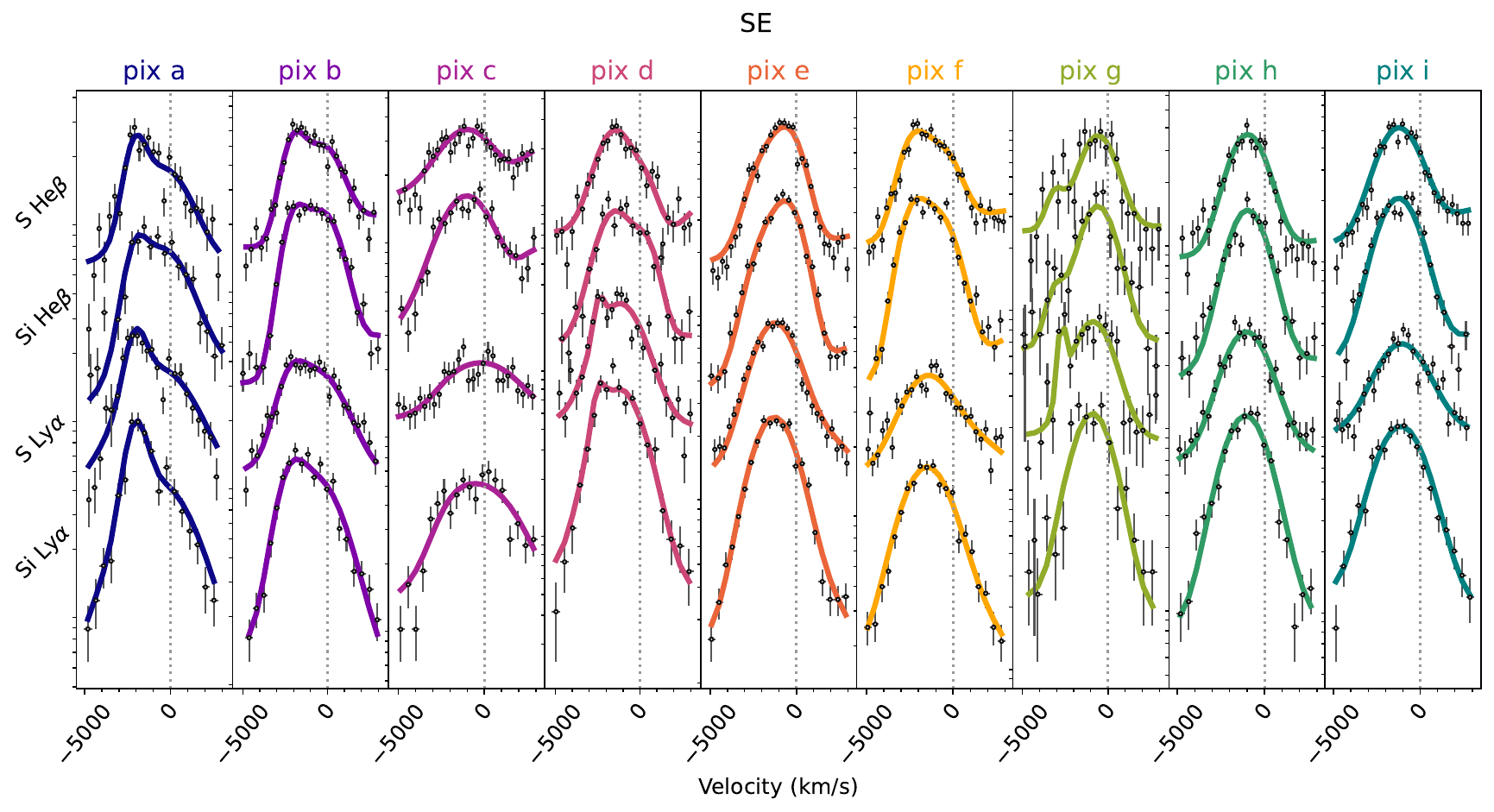}
\includegraphics[width=\textwidth]{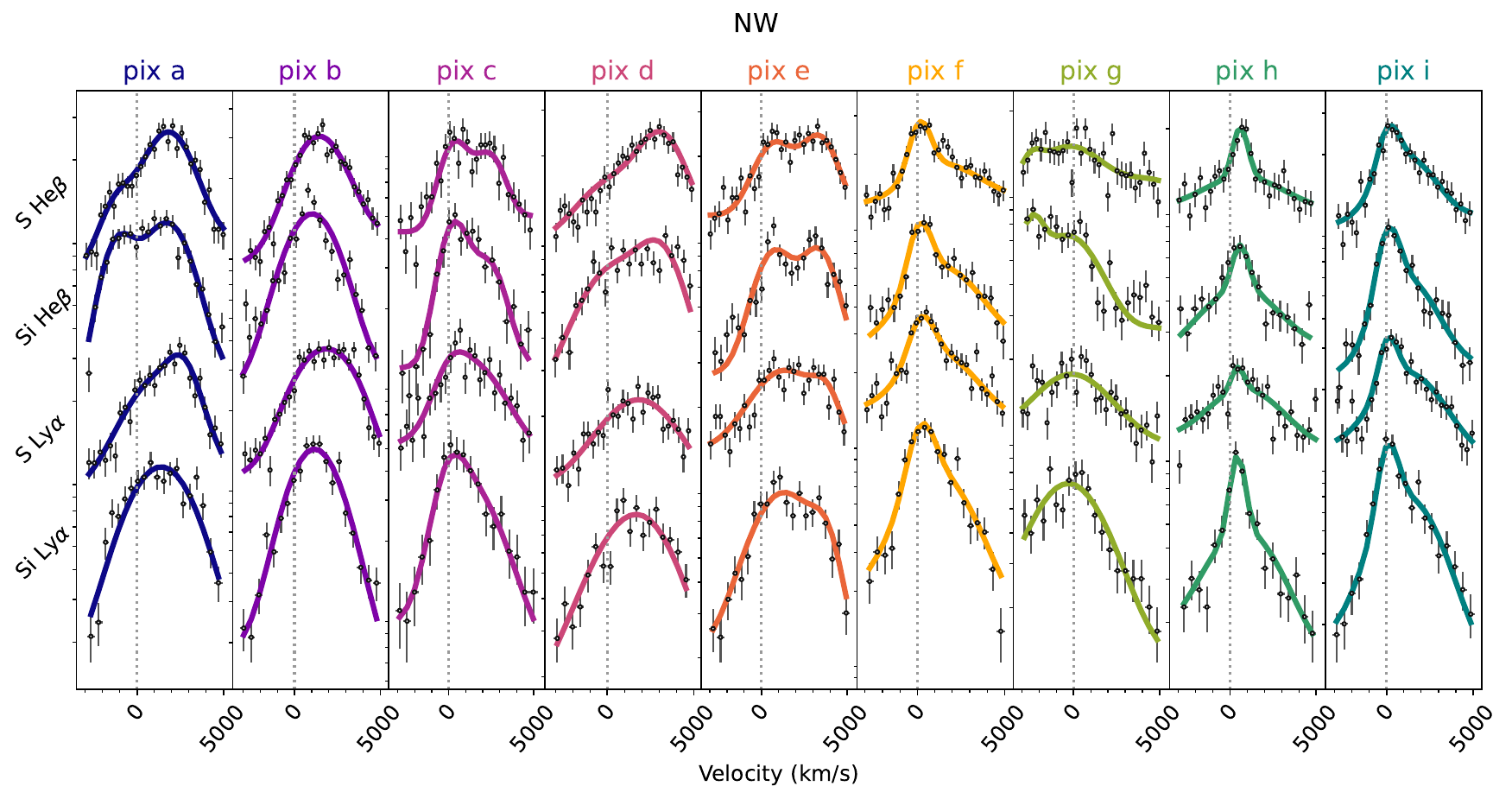}
    \caption{\label{fig:lineprofiles_kms}
    Alternative figure to Fig.~\ref{fig:lineprofs}, but now the line shapes are shown as a function of radial velocity rather than photon energy. Each column represents a ``super pixel", and shows the S He$\beta$, Si He$\beta$, S Ly$\alpha$, and Si Ly$\alpha$ lines, respectively, scaled for display purposes.
    Note that the He-$\beta$ profiles are distorted due to the presence of Li-like satellite lines.
    Alt text: Line profiles of S He$\beta$, Si He$\beta$, S Ly$\alpha$, and Si Ly$\alpha$, in velocity space are shown vertically next to one another for easier comparison of line profiles.}
\end{figure*}

We show here Fig.~\ref{fig:lineprofiles_kms}, which is an alternative for Fig.~\ref{fig:lineprofs}, but showing the line profile as
a function of radial velocity.

%\bibliographystyle{aasjournal}

% Journals(e.g. A\&A,ApJ,AJ,NMRAS,PASP ...)
% Authors, Year, Journal, Vol#, Page# 
\end{document}